\documentclass[fleqn,usenatbib]{mnras}

\usepackage{times} 
\usepackage{listings}
\usepackage{graphics}
\usepackage{graphicx,xspace}
\usepackage{amsmath}
\usepackage{amssymb}
\usepackage{hyperref}
\usepackage{lscape}
\usepackage{rotating}
\usepackage{placeins}
\usepackage{natbib}
\usepackage{color}
\usepackage{txfonts}
\usepackage{mathptmx}
\usepackage{newtxtext,newtxmath}

\usepackage{subcaption}
\usepackage[dvipsnames]{xcolor}
\usepackage{caption}

\usepackage{grffile}
\usepackage{tabularx}
\usepackage{adjustbox}
\usepackage{array, makecell, cellspace}
\usepackage{multirow}
\usepackage{siunitx} 
\usepackage{tikz}

\usepackage{multicol}
\usepackage{changepage}
\usepackage[autostyle, english = american]{csquotes}

\newcommand{\hi}{H{\sc\,i}\xspace}

\newcommand{\perbeam}{beam$^{-1}$}

\usepackage{amsmath}

\title[Deep \hi survey of COSMOS with FAST]{Deep extragalactic \hi survey of the COSMOS field with FAST}

\author[H. Pan et al.]{Hengxing Pan$^{1}$\thanks{E-mail:hengxing.pan@physics.ox.ac.uk}, Matt J.~Jarvis$^{1,2}$, Ming Zhu$^{3,4}$, Yin-Zhe Ma$^{5}$, Mario G. Santos$^{2,6}$, Anastasia A. Ponomareva$^{1}$,\\ 
\newauthor
Ian Heywood$^{1,7,6}$, Yingjie Jing$^{4}$, Chen Xu$^{4,8}$, Ziming Liu$^{4,8}$, Yogesh Chandola$^{9}$ and Yipeng Jing$^{10,11}$\\
\\
$^{1}$Astrophysics, University of Oxford, Denys Wilkinson Buiding, Keble Road, Oxford OX1 3RH, UK\\
$^{2}$Department of Physics and Astronomy, University of the Western Cape, Cape Town 7535, South Africa\\
$^{3}$Guizhou Radio Astronomical Observatory, Guizhou University, Guiyang 550000, People’s Republic of China\\
$^{4}$National Astronomical Observatories, Chinese Academy of Sciences, Beijing 100101, People’s Republic of China\\
$^{5}$Department of Physics, Stellenbosch University, Matieland 7602, South Africa\\
$^{6}$South African Radio Astronomy Observatory (SARAO), 2 Fir Street, Observatory, 7925, South Africa\\
$^{7}$Centre for Radio Astronomy Techniques and Technologies, Department of Physics and Electronics, Rhodes University, PO Box 94, Makhanda 6140, South Africa\\
$^{8}$School of Astronomy and Space Science, University of Chinese Academy of Sciences, Beijing 100049, China\\
$^{9}$Purple Mountain Observatory, Chinese Academy of Sciences, 10, Yuan Hua Road, Qixia District,  Nanjing, 210023, China\\
$^{10}$Department of Astronomy, School of Physics and Astronomy, Shanghai Jiao Tong University, Shanghai, 200240, People’s Republic of China\\
$^{11}$Tsung-Dao Lee Institute, and Shanghai Key Laboratory for Particle Physics and Cosmology, \\
$\hspace{2mm}$Shanghai Jiao Tong University, Shanghai, 200240, People’s Republic of China\\
}
\date{Accepted XXX. Received YYY; in original form ZZZ}

\pubyear{2024}

\begin{document}
\label{firstpage}
\pagerange{\pageref{firstpage}--\pageref{lastpage}}
\maketitle

\begin{abstract}

We present a deep \hi survey at L-band conducted with the Five-hundred-meter Aperture Spherical radio Telescope (FAST) over the COSMOS field. This survey is strategically designed to overlap with the MIGHTEE COSMOS field, aiming to combine the sensitivity of the FAST and high-resolution of the MeerKAT. We observed the field with FAST for approximately 11 hours covering $\sim2$ square degrees, and reduced the raw data to \hi spectral cubes over the frequency range 1310-1420 MHz. The FAST-\hi data reach a median 3$\sigma$ column density of $N_{\rm{HI}} \sim 2\times 10^{17}$ cm$^{-2}$ over a $\sim5\,{\rm km}\,{\rm s}^{-1}$ channel width, allowing for studies of the distribution of \hi gas in various environments, such as in galaxies, the Circum-Galactic Medium (CGM) and Intergalactic Medium (IGM). We visually searched the spectral cubes for \hi sources, and found a total of 80 \hi detections, of which 56 have been cross-matched with the MIGHTEE-\hi catalogue. With the cross-matched sources, we compare their \hi masses and find that the total \hi mass fraction in the IGM and CGM surrounding the galaxy pairs is statistically higher than the \hi fraction surrounding the isolated galaxies by a difference of $\sim$13$\pm4$\%, indicating that the CGM and IGM associated with interacting systems are richer in neutral hydrogen compared to those around isolated galaxies in the local Universe. We also describe several FAST-MeerKAT synergy projects, highlighting the full potential of exploiting both single-dish and interferometric observations to study the distribution and evolution of the diffuse \hi gas.

\end{abstract}




\section{Introduction}

Galaxy growth relies on the continuous gas accretion from the surrounding environments. Cold gas within galaxies in the form of neutral hydrogen (\hi) serves as the reservoir from which molecules and stars subsequently form. The Intergalactic Medium (IGM) connects galaxies and acts as a route to fueling them, and the Circum-Galactic Medium (CGM) traces the flow of matter, energy, and enriched elements between galaxies and the IGM \citep[e.g.][]{Tumlinson_2017, Martin2019,Saintonge_2022,Decataldo2024}. Understanding the distribution of baryons within galaxies, the CGM and IGM is therefore a fundamental problem in astrophysics, with critical implications for the growth mechanisms of galaxies \citep[e.g.][]{Sancisi_2008,Putman_2017,Faucher2023} and can help solve the long-lasting ``missing" baryon problem \citep{Shull_2012, Ma_2015, Macquart_2020, Li2024}.

In the past few decades, several \hi galaxy surveys with large sky coverage have been undertaken by radio telescopes,  such as the \hi Parkes All-Sky (HIPASS) Survey  \citep{barnes2001h} and the Arecibo Legacy Fast ALFA (ALFALFA) survey \citep{giovanelli2005arecibo}, to detect the 21-cm emission line from neutral hydrogen in galaxies. However, these surveys are constrained to the local Universe ($z\lesssim0.06$), due to the limited sensitivity and frequency range of the telescopes. On the other hand, our view of the IGM and CGM is based largely on the powerful but restricted information from QSO absorption line surveys that probe the Lyman-$\alpha$ forest \citep[e.g.][]{Becker_2011,Keating_2013,Telikova_2019}. Unfortunately, emission from intergalactic baryons is difficult to observe because of current telescope sensitivities, which limit studies to column densities $N_{\rm{HI}} \gtrsim 10^{19}$ (atoms) cm$^{-2}$, which is the realm of Damped Ly$\alpha$ (DLA) systems and sub-DLAs. Below column densities of $\sim 10^{19}$ cm$^{-2}$, the neutral fraction of hydrogen decreases rapidly due to the transition from optically-thick to optically-thin gas ionized by the metagalactic ultraviolet radiation. However, below $\sim 10^{18}$ cm$^{-2}$ the gas is optically thin and the decline in neutral fraction is much more gradual \citep{Popping_2009, popping2015observations}. This gas, generally thought to be residing in filamentary structures, serves as the the pristine reservoir capable of fueling future star formation in galaxies, and could provide a direct signature of the predicted smooth cold-mode accretion to dominate gas acquisition in star-forming
galaxies today \citep[]{Kere__2009,deblok2017overview}. Hence, exploring the $N_{\rm{HI}} < 10^{18}$ cm$^{-2}$ regime is crucial to deepen our understanding of the reservoir of baryons that shape galaxy formation and evolution.

With the sensitivity of the Five-hundred-meter Aperture Spherical Radio Telescope \citep[FAST;][]{NAN_2011}, we now have a facility with the potential to detect these faint \hi emissions within a reasonable integration time, as indicated, for example, by the FAST All Sky HI survey \citep[FASHI;][]{zhang2024}. However, due to the large beam size ($\sim$3 arcmin) of FAST, disentangling the IGM/CGM from the \hi gas within individual galaxies is limited to the very local nearby Universe. Fortunately, with the excellent angular resolution ($\sim$10
arcsec)  of the MeerKAT radio telescope \citep{Jonas2016mksJ}, we can identify the \hi within these galaxies directly, and then use a combination of data from FAST and MeerKAT to extract the IGM/CGM signal at the $N_{\rm{HI}} < 10^{18}$ cm$^{-2}$ regime thoroughly. To achieve this, we carried out a series of FAST observations in L-band across the COSMOS field, overlapping with one of the MeerKAT International GHz Tiered Extragalactic Exploration \citep[MIGHTEE;][]{Jarvis2016} fields.

In this paper, we introduce our deep \hi survey aimed at detecting the \hi gas not only in the galaxies but also in the IGM and CGM with FAST. We describe the data processing for FAST and the ancillary data in Section~\ref{sec:data}. We then present the FAST-\hi detections, and investigate the $M_{\rm HI}-M_{\star}$ relation between the FAST-\hi and the MIGHTEE-\hi samples in Section \ref{sec:results} ending with a list of FAST-MeerKAT synergy projects, and then conclude in Section \ref{sec:conclusions}. We use the standard $\Lambda$CDM cosmology with a Hubble constant $H_{0}=67.4$ ${\rm km}\,{\rm s}^{-1}\,{\rm Mpc}^{-1}$, total matter density $\Omega_{\rm m}=0.315$ and dark energy density $\Omega_{\Lambda}=0.685$ \citep{planck2020} in our analysis.

\section{Data}
\label{sec:data}

\subsection{FAST-\hi}

\begin{figure}
    \centering
    \includegraphics[width=0.9\columnwidth]{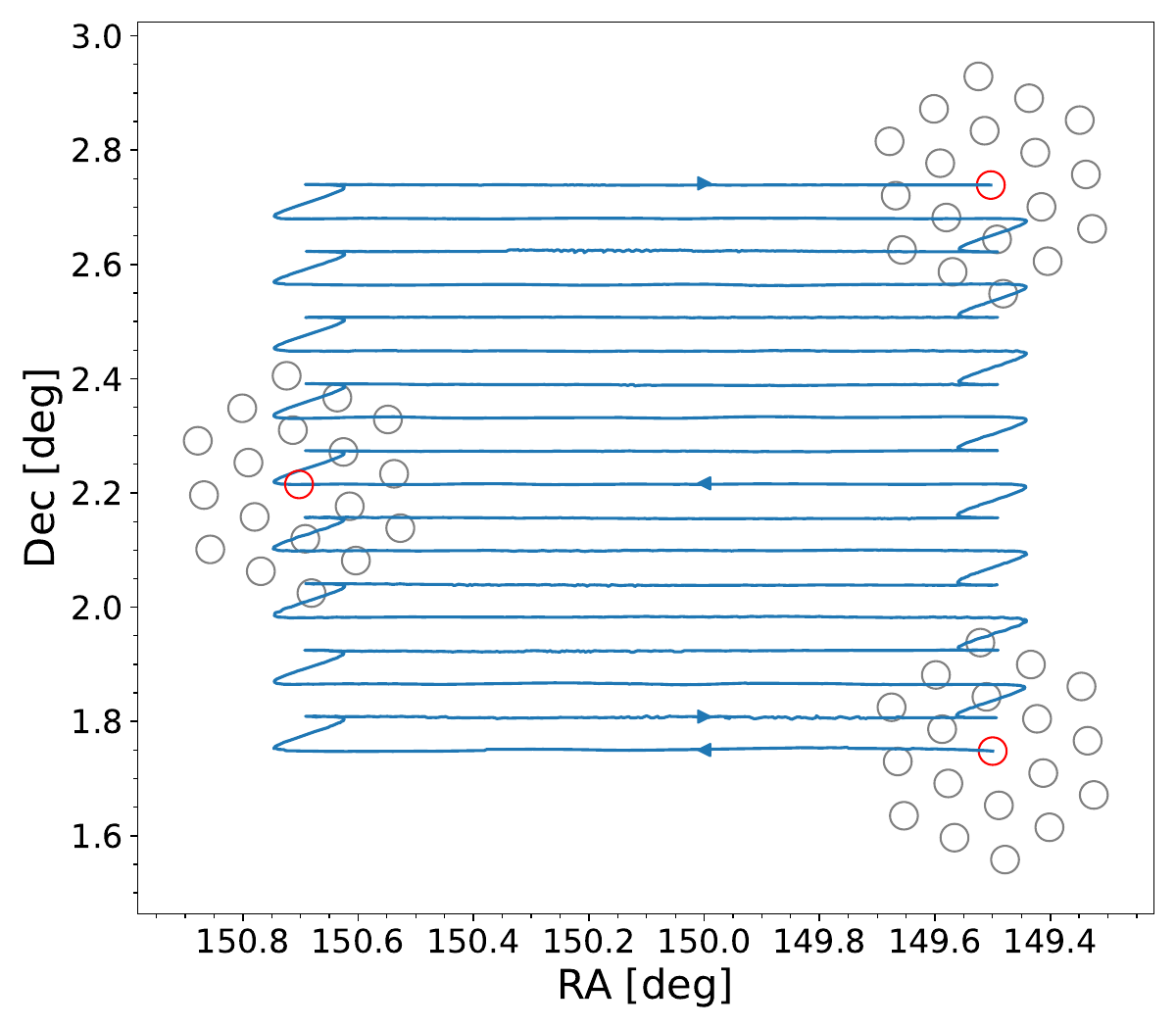}
  \caption{Footprints of the FAST 19 beams in the COSMOS field. Each beam size is 2.9$^\prime$ at $z=0$. The arrows and the blue line show the direction of the beam 01 (i.e. the centre beam denoted by a red circle) scanning the field with the MultibeamOTF mode.}
  \label{fig:field}
\end{figure}

FAST has a 500-meter diameter dish constructed in a natural depression in Guizhou province of China and is the largest single-dish telescope in the world. We made observations between the 1st and 6th of January 2022 for six nights over $\sim$2 square degrees of the COSMOS field with FAST's MultiBeamOTF (On-The-Fly) mode for a total amount of 11-hours integration time on source. The footprints are shown in Figure~\ref{fig:field}. The 19 beams are rotated clockwise by a angle of 23.4$^\circ$ to ensure a uniform coverage of the field in this horizontal scanning mode. The scan gap is fixed at 3.5 arcmins to reduce the frequency of the source changing. The flux calibrator 3C237 was observed on the first day of the observations with FAST's MultiBeamCalibration mode. The observations are listed in Table~\ref{obs_history}.

\begin{table*}
\centering
\caption{Observation log for the COSMOS field and the calibrator 3C237.}
\begin{tabular}{lllll}
\hline
\hline
Name	& Mode	& Length(s)	& Start Time &	End Time \\
\hline
COSMOS	& MultiBeamOTF	&       6642	&2022-01-06 02:50:00	&2022-01-06 04:40:42 \\	 
COSMOS	& MultiBeamOTF	&       6642	&2022-01-05 02:50:00	&2022-01-05 04:40:42 \\	 
COSMOS	& MultiBeamOTF	&       6642	&2022-01-04 02:50:00	&2022-01-04 04:40:42 \\	 
COSMOS	& MultiBeamOTF	&       6642	&2022-01-03 03:12:00	&2022-01-03 05:02:42 \\	 
COSMOS	& MultiBeamOTF	&       6642	&2022-01-02 03:00:00	&2022-01-02 04:50:42 \\	 
3C237	& MultiBeamCalibration &2700	&2022-01-01 05:06:00	&2022-01-01 05:51:00 \\	 
COSMOS	& MultiBeamOTF	&       6642	&2022-01-01 03:05:00	&2022-01-01 04:55:42 \\
\hline
\end{tabular}
\label{obs_history}
\end{table*}
\hfill

We use the \texttt{HiFAST}\footnote{\url{https://hifast.readthedocs.io/en/v1.3}} pipeline \citep{Jing2024} to reduce the raw spectral data recorded by the 19-beam L-band receiver of FAST. This pipeline is a dedicated, modular, and self-contained calibration and imaging system designed for processing the \hi data from FAST. The modules in the pipeline can be combined as needed to process the data from various FAST observation modes, including tracking, drift scanning, On-The-Fly mapping, and most of their variants. We provide information on a few key modules in the following subsections. 

\subsubsection{Calibration}
We first calibrate the antenna temperature with a standard 10\,K noise diode, and the noise source was injected every 32 seconds. However, the gain fluctuations depend on the condition of the telescope and the receiver during observation. Therefore, we observed the calibrator 3C237 during the initial stage of observing the target sources to calibrate the flux density scale. We use the latest continuum measurements of the calibrator 3C237 as the model across the frequency range of 900-1670 MHz from MeerKAT. The flux density as a function of frequency is described by a power law,
\begin{equation}
    S(v) = S_{\rm 1.4GHz}\left(\frac{v}{\rm 1.4\;GHz}\right)^\alpha,
	\label{eq:cal}
\end{equation}
where $S_{\rm 1.4GHz}=6.533 \pm 0.147$ Jy, and $\alpha=-0.919 \pm 0.01$ \citep{taylor2021}. The uncertainty on the flux density is $\sim$2 percent at the frequency of $\sim$1.4 GHz, which is taken into account when the error of the \hi flux is determined in addition to the effect of the thermal noise. The data processing of the calibrator observation is detailed by \cite{Liu_2024}.

\subsubsection{Baseline subtraction}
We then fit the spectral baseline with two approaches: 1) the asymmetrically reweighted penalized least squares smoothing (arPLS) method \citep{Baek2015} and 2) the minimum of the medians (MinMed) method \citep{Putman_2002}. The arPLS method iteratively refines the baseline by adjusting and applying weights to the data points until the weights converge. The MinMed method splits the scanned spectra into several parts and each part comprises several segments along the time axis for each channel, and then the minimum of the medians in each segment is taken as the reference (off-source) spectra which is subtracted in that part afterwards. These two approaches complement each other to account for continuum variations across a broad range of the frequency scales.

\subsubsection{Standing wave removal}
Standing waves arise because the radio signal enters the receiving system by two paths of different length due to the present design of radio telescopes. The signal taking the longer path suffers a delay, and then correlates with the signal in the shorter path as a function of time. To obtain the power spectrum, the correlation function is Fourier-transformed to a sinusoidal variation across the frequency band. The difference of the two path lengths is $\sim$276 meters for FAST, which corresponds to a time delay of $\sim$0.92 microseconds, therefore the resulting standing wave has a period of $\sim$1.09 MHz in frequency space. Hence we fit the standing wave with a Fourier approach (Xu et al. in prep). This approach involves applying a Fourier transform to the  ``waterfall data'' (i.e. spectral data as a function of time and frequency) in the frequency space, transforming them into the delay space. We then select the peak amplitude and its surrounding modes for performing an inverse Fourier transform at each timestamp. The standing wave is modelled in this way, and then removed from the baseline subtracted spectra.

\subsubsection{RFI flagging}
The radio frequency interference (RFI) often refers to artificial radio emission from human-made electronics such as satellites and civil aircraft. It corrupts astronomical measurements and cannot be well modelled, therefore has to be masked out. The RFI presented in our data can be loosely classified as three types: the time domain RFI that cover a wide frequency band from geosynchronous satellites, the narrow band single channel RFI, and polarized RFI. We identify the time domain RFI by averaging the 2-dimensional waterfall data along the frequency axis, and select the narrow band single channel RFI by averaging the waterfall data along the time axis. The polarized RFI is flagged as the \hi signal tends to be non-polarized. We also perform a visual checkup to manually mask out the RFI that is not well defined by the above types. Overall, 10-20 percent of the data is masked out across the bandwidth of 1310-1420 MHz, with most of the affected data due to the wide band RFI. The RFI contamination at the lower frequency ($<$1310 MHz) band is complicated \citep{Zhang_2022} and we leave it for a separate paper to describe the flagging procedures. 

\subsubsection{Gridding and smoothing}
We correct the coordinate frame for the Doppler effect due to the motion of the earth, and finally grid the individual spectra into a 3-dimensional cube with a weighting scheme that accounts for the distance of each spectrum from the grid point \citep{Mangum_2007}. The channel width of raw FAST data is 7.63 kHz, and we smooth the \hi data cube with a Hanning window along the frequency to a resolution of 22.9 kHz which is $\sim$$4.83\,{\rm km}\,{\rm s}^{-1}$ at $z=0$. At this stage, the data cube still has a low-level continuum residual, which has been further subtracted by fitting a second-degree polynomial function to the residual per pixel after applying a 3 sigma-clipping. We note that a \hi cube  with a lower-velocity resolution of 26 ${\rm km}\,{\rm s}^{-1}$ is also made for the visual source finding only.

\begin{figure*}
\centering
    \includegraphics[width=0.8\textwidth]{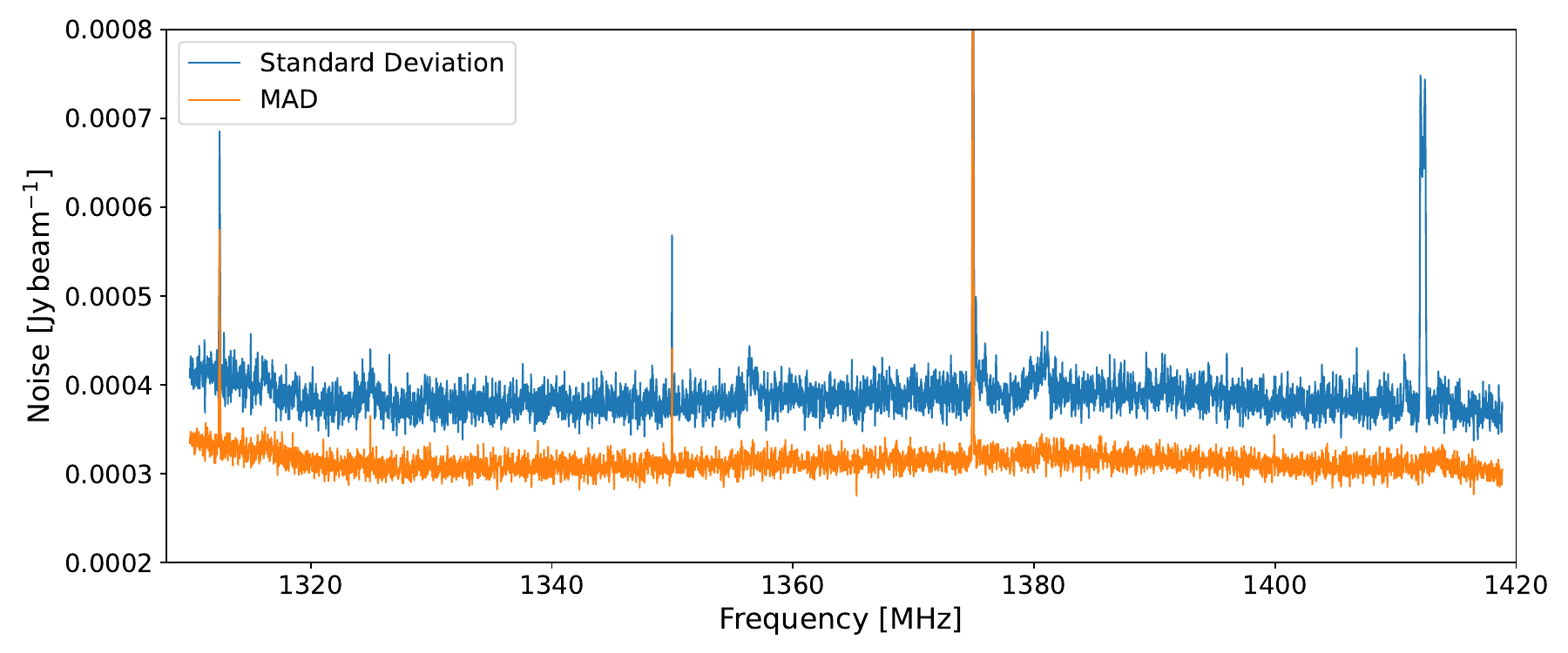}
  \caption{Noise vs frequency for the \hi cube with a channel width of $\sim$23 kHz. The blue and orange lines are the noise levels estimated by using the standard deviation and median absolute deviation (MAD) for each channel, respectively. The pronounced noise spikes are due to the flagging of the single channel RFI.}
  \label{fig:rmsvsfreq}
\end{figure*}

\begin{figure}
    \centering
    \includegraphics[width=\columnwidth]{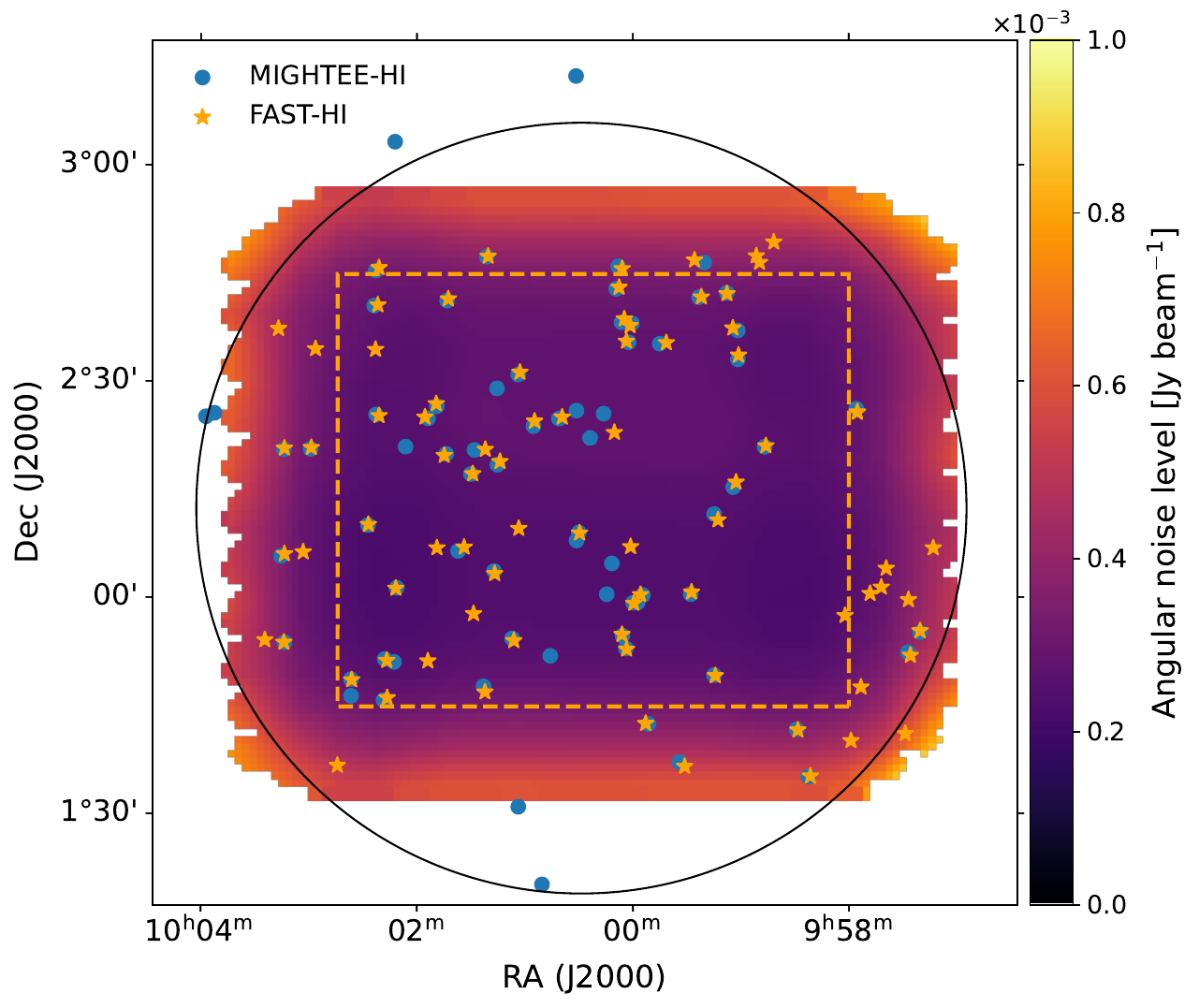}
  \caption{FAST observation coverage (in color-coded area) overlapped over MIGHTEE Early Science COSMOS field (in black circle). The blue and orange stars are the MIGHTEE-\hi and FAST-\hi detections, respectively. The colour scheme indicates the noise level on a projected 2-dimensional sky. The dashed orange box delineates the perimeter within which the centre beam of the 19 beams has passed over, and the noise level increases gradually towards the boundaries of the whole field due to limited number of beams available. The black circle shows the MeerKAT primary beam at $\sim$10\% level.}
  \label{fig:spacial}
\end{figure}

\subsubsection{Data cube}
We show the RMS against frequency in Figure~\ref{fig:rmsvsfreq}, and show that our final \hi cube has a median channel noise of $\sim$385 $\mu$Jy \perbeam, which roughly meets our expectation considering that the COSMOS field is close to the equator where the aperture efficiency of FAST is reduced by $\sim$10\% with a zenith angle of $\sim30^{\circ}$  \citep{Jiang_2020}, and the additional effects of flagging the RFI. The noise level is relatively constant across the whole band between 1310 and 1420 MHz with a few pronounced spikes due to the flagging of the narrow band single channel RFI. However, the values from the median absolute deviation (MAD) is $\sim$20\% lower than the standard deviation. This means that the noise does not follow a perfect Gaussian distribution in the spatial domain as we can see from the color-coded noise map in Figure~\ref{fig:spacial}, although the areas outside the dashed orange box are excluded. Indeed, our observations suffered from the strong wide band RFI from a Geosynchronous satellite and lost about half of the integration time in the upper spectral coverage of the COSMOS field, likely leading to suboptimal continuum subtractions and a moderate level of non-Gaussian noise behaviour. We provide a few key parameters for our FAST data in Table~\ref{fast_data}.

\begin{table}
\caption{Key observational characteristics of the FAST-\hi COSMOS data. Note that the total sky coverage is $\sim$2 deg$^2$ where the centre sensitive area is $\sim$1.2 deg$^2$.}
\label{fast_data}
\centering
\begin{tabular}{lc}
\hline
\hline
Key parameters & FAST \\
\hline
Centre coordinate & 10h00m, +02d14m \\
Area covered    & $\sim$2 deg$^2$ \\
Obs. time & 11 hours \\
Frequency range & $1310-1420$ MHz \\
Channel width   & 22.9 kHz  \\
Pixel size      & 60$^{\prime\prime}$                                        \\
Median \hi channel rms noise & $\sim$385 $\mu$Jy \perbeam              \\
Beam  & 174$^{\prime\prime}$ $\times$ 174$^{\prime\prime}$ ($z=0$)                     \\
$N_\mathrm{HI}$ sensitivity ($3\sigma$)  & $\sim$2$\times10^{17}$$\mathrm{cm}^{-2}$   \\
\hline
\end{tabular}
\end{table}

\subsection{Ancillary data}

The COSMOS field is covered by various multi-wavelength photometric and spectroscopic surveys ranging from X-ray to radio
bands. In particular, we exploit the MIGHTEE-\hi data which is the \hi emission project within the MIGHTEE survey undertaken by the interferometric MeerKAT telescope as one of eight large survey projects \citep{Jarvis2016, Maddox_2021}.
We make use of the \hi galaxy catalogue compiled from the MIGHTEE-\hi Early Science data which are collected with the MeerKAT 4k correlator mode in the L-band with a frequency range from 900 to 1670 MHz \citep{Ponomareva_2023}. 

However, we use the latest MIGHTEE Data Release 1 which is observed with the 32k channels \citep{Heywood_2024} for comparing the \hi correlator mode.
We smooth a cubelet surrounding each source in the Early Science catalogue provided by \cite{Ponomareva_2023}, and clip it at a 3$\sigma$ threshold as a mask for removing the
noise following \citep{Ponomareva_2021}. 
The MeerKAT 32k correlator mode has 32,768 channels with a spectral resolution of 26.1 kHz (i.e. 5.5 kms$^{-1}$ at 1420 MHz). The COSMOS field was observed in this mode for a total of 15$\times$8h tracks in a tightly-dithered mosaic that spans $\sim$2 deg$^2$, and each pointing is imaged with robustness parameters of 0.0 and 0.5 \citep{briggs1995american} following up on two rounds of RFI flagging, self-calibration, and the visibility-domain continuum subtraction. Then all pointings are brought together for homogenisation, mosaicking, and a further image-domain continuum subtraction \citep[][]{Heywood_2024, Jarvis2024}. We use the data product with the robustness of 0.5 (i.e. angular resolution $\sim$16 arcseconds across 1300-1420 MHz) to have a well balanced combination of the image resolution and sensitivity for detecting the \hi gas within galaxies. We note that a robust catalogue is currently not available for the 32k data, which is why we use the Early Science catalogue to match to the galaxy positions.

The stellar masses of the \hi galaxies are derived using the Spectral Energy Distribution (SED) fitting code LePhare \citep{Ilbert2006}, and the uncertainty on the stellar mass is conservatively assumed to be $\sim$0.1 dex, due to assumptions made in the SED fitting process on star formation history and initial mass function etc. \citep{Adams_2021}. The optical spectroscopic data are mainly taken from the Deep Extragalactic VIsible Legacy Survey (DEVILS) survey \citep{Davies_2018,Hashemizadeh_2021} for cross-matching the optical counterparts with our FAST-\hi detections. The ALFALFA \hi data are from \cite{Haynes_2018}.

\section{Results}
\label{sec:results}

\subsection{FAST-\hi source finding}

\begin{figure*}
    \centering
\begin{subfigure}[c]{0.45\textwidth}
    \includegraphics[width=0.95\columnwidth]{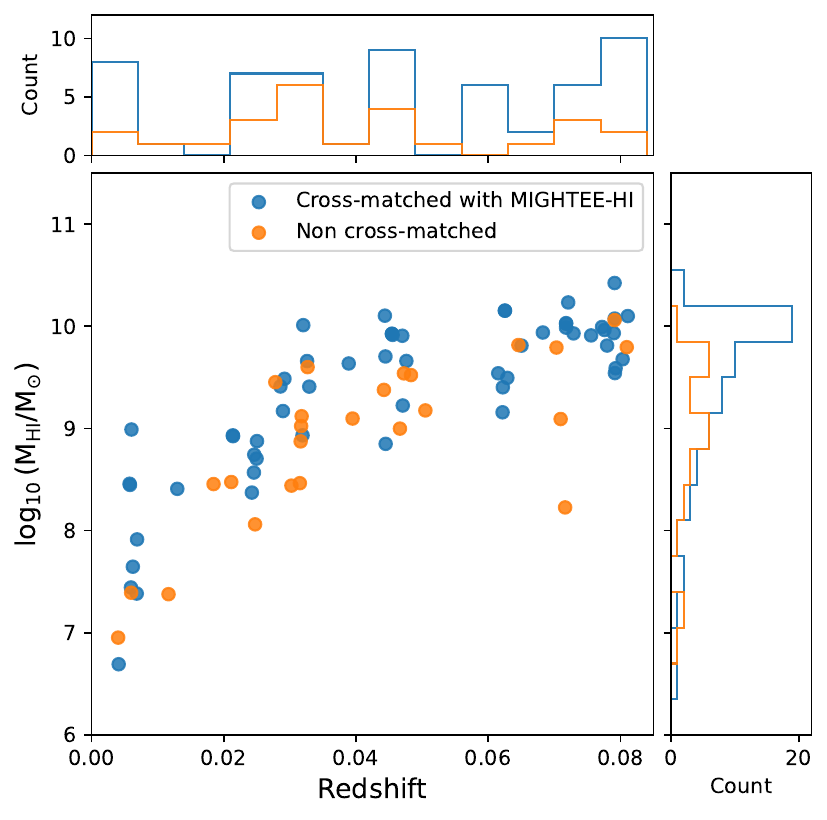}
\end{subfigure}
\begin{subfigure}[c]{0.45\textwidth}
    \includegraphics[width=0.95\columnwidth]{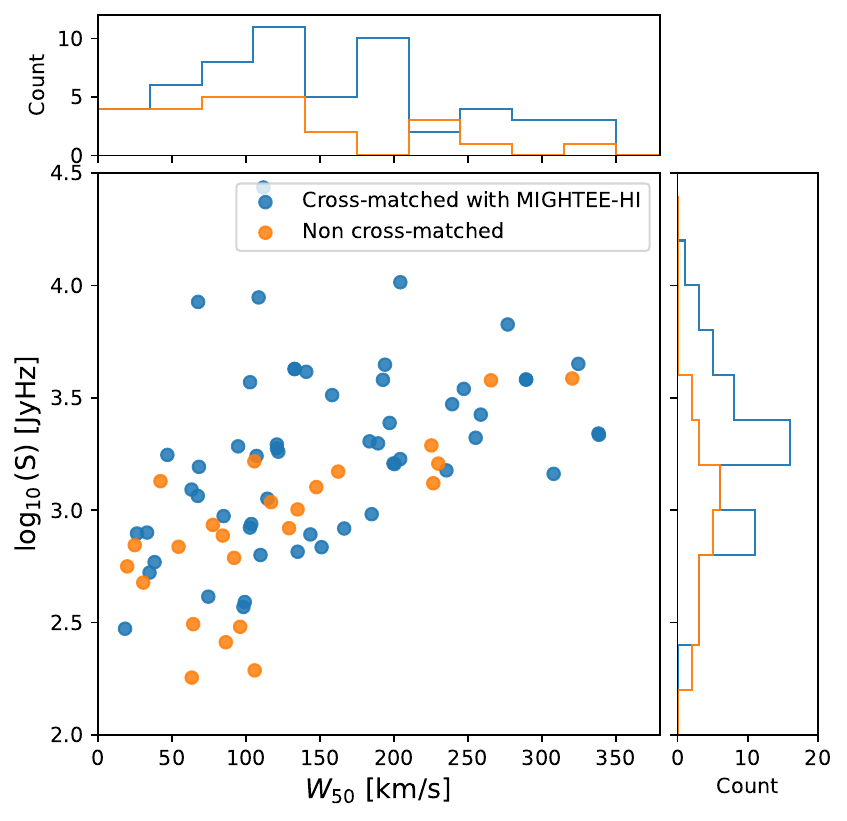}
\end{subfigure}
  \caption{FAST-\hi mass as a function of redshift  (left) and FAST-\hi flux against the velocity width $W_{50}$ (right) for the cross-matched (blue) and non cross-matched (orange) samples between FAST- and MIGHTEE-\hi detections.}
  \label{fig:xmatched}
\end{figure*}

We find a total of 80 sources after eyeballing the \hi images with the Cube Analysis and Rendering Tool for Astronomy \citep[CARTA;][]{Comrie2021}, and then employ SoFiA 2 \citep{Serra_2015,Westmeier_2021} to refine the source finding and parameterise the \hi source in subcubes centered at the visually-identified positions. The subcubes have an angular size ranging from 3 to 10 beam sizes (i.e. $\sim$9-30 arcmins) based on the source size. SoFiA 2 is a fully automated 3D source finding pipeline for extragalactic \hi surveys. It first convolves the \hi cube or subcube with various smoothing
kernels, and selects voxels with absolute values above a predefined threshold, then links the selected voxels together if they are close enough, to form the mask of the source. For our FAST data, the flux detection threshold is 5$\sigma$, where $\sigma$ is the local noise
level. We chose this relatively high threshold to mitigate the effect of generally less stable baselines of single-dish data and demonstrate that we can recover the intrinsic fluxes with a median difference fraction of $\sim$1 percent with injected fake sources (see  Appendix~\ref{sec:injection}). The kernel has an angular size of 0, 3 and 6 pixels (i.e. $\sim$2 beam sizes maximum as each pixel size is $\ang{;1;}$) in the projected sky direction, and a velocity width of 0, 3, 7, 15, and 31 channels (i.e. 155 ${\rm km}\,{\rm s}^{-1}$ maximum). 

For each source, the integrated flux $S$ is measured by summing the flux density values of all pixels contained in the source mask, and multiplying by the spectral channel width, and then dividing by the beam solid angle. The uncertainty of the integrated flux is estimated by assuming Gaussian error propagation of the local noise and correcting for the correlation of the spatial pixels due to the finite beam size. The \hi mass is  determined under the optically thin gas assumption by 
\begin{equation}
    \frac{M_{\rm HI}}{M_{\odot}} \simeq 49.7\left(\frac{D_L}{\rm Mpc}\right)^2 \left(\frac{S}{\rm JyHz}\right),
	\label{eq:factor}
\end{equation}
where $M_{\rm HI}$ is the \hi mass, and $D_L$ is the luminosity distance \citep{meyer2017tracing}.

The line width $W_{50}$ of the integrated spectral profile is measured by moving inwards from both ends of the spectrum until the signal exceeds 50\% of the peak flux density in the spectrum, and the error of the $W_{50}$ is typically less than 10 percent for sources with the SNR$>$10, and 5 percent is adopted for our data if the SNR is greater than 20 \citep{Westmeier_2021}.

\subsection{FAST vs MeerKAT}

\subsubsection{Cross matching}

After cross-matching the sample of 80 FAST-\hi detections with the MIGHTEE-\hi Early Science catalogue \citep[][]{Ponomareva_2023}, but using the measurements from Date Release 1 \citep{Heywood_2024}, by setting the maximum angular separation of 3 arcminites and velocity offset of $\sim$200 ${\rm km}\,{\rm s}^{-1}$, we find 56 cross-matched sources. A comparison of their spatial distribution is shown in Figure~\ref{fig:spacial}. The rate of finding a MIGHTEE-\hi counterpart is clearly higher in the centre region with the highest sensitivity.  We plot their \hi masses against redshift in the left panel of Figure~\ref{fig:xmatched}, and the \hi flux against $W_{50}$ in the right panel to investigate the differences between the FAST-\hi and MIGHTEE-\hi detections. It shows that the non MIGHTEE-detected sources are dominated by the low-flux or narrow-line width \hi galaxies as would be expected given the coarse spectrum resolution of 44 ${\rm km}\,{\rm s}^{-1}$ for the MIGHTEE-\hi Early Science data and the limited sensitivity. However, there are a few non cross-matched \hi sources that have a moderate amount of \hi gas and a velocity width larger than 150 ${\rm km}\,{\rm s}^{-1}$ in the FAST data. In other words, some non cross-matched \hi sources appear to be relatively bright galaxies, and are detected in FAST-\hi but not in the MIGHTEE-\hi 4k catalogue. This difference is likely indicating the fundamental difference between MeerKAT and FAST as an interferometer is generally less sensitive to extended/diffuse emission while a single dish telescope is sensitive to both compact and extended emission. The source finding with the MIGHTEE-\hi 32k data will help further examine these sources. There are also some sources that are detected by MeerKAT but not FAST from our initial visual source finding. The presence of systematics in the FAST data complicates baseline subtraction, potentially contributing to the ``missing sources'' from the FAST data, especially for the faint ones. By using the positions from the MIGHTEE-\hi detections, we extract the FAST-\hi fluxes for the missing ones and find that these non-detections are indeed very faint with signal to noise ratios less than $\sim10$. Their spectra and masses are shown in Appendix~\ref{sec:non-det}.

\begin{figure}
    \centering
    \includegraphics[width=\columnwidth]{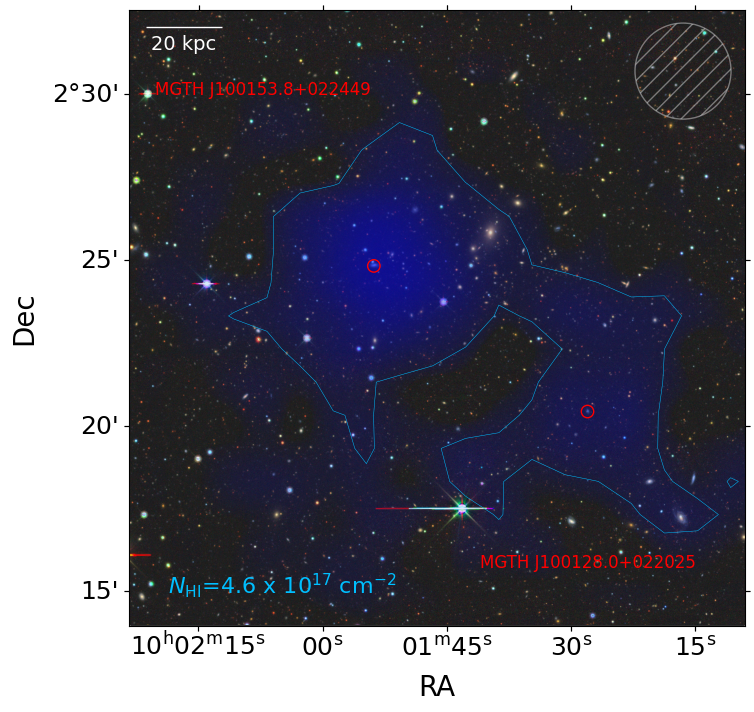}
  \caption{Example of the FAST-\hi  detection, with the blue halo of the interpolated \hi moment-0 map overlaid on DECaLS \protect\citep{Dey_2019} false color image, where the contour (light blue line) with a column density of 4.6$\times 10^{17}$\,$\mathrm{cm}^{-2}$ is $\sim$1.5 times the local noise level on the moment-0 map. The red circles denote the MIGHTEE-\hi detections with their names labelled. The hatched grey circle on the upper right corner shows the FAST beam size of 2.9 arcmins.}
  \label{fig:example}
\end{figure}

The cross-matched detections allow us to study the \hi galaxies and their surrounding CGM and IGM in detail by comparing the FAST and MIGHTEE-\hi images, and also further help to investigate the Tully-Fisher relation, $M_{\rm HI}$-size relation, $M_{\rm HI}$-$M_{\star}$ relation, and the \hi mass function that have been investigated separately by the MIGHTEE-\hi team \citep[e.g.][]{Ponomareva_2021,Ponomareva_2023,Rajohnson_2022,Pan_2023}. An example of a FAST-\hi detection is shown in Figure~\ref{fig:example}, where a thin \hi bridge appears to be between the two nearby galaxies at $z\sim0.0068$ with  a column density of $N_{\rm HI}< 10^{18}$\,$\mathrm{cm}^{-2}$ possibly due to the tidal interaction given their close distance of $\sim100$ kpc. However, we note that the lowest contour is $\sim$1.5 times the local noise level on the moment-0 map.

The red circles are the MIGHTEE-\hi counterparts J100153.8+022449 and J100128.0+022025 with $\log_{10}(M_{\rm HI}/M_\odot)=7.66\pm0.02$ and 7.23$\pm0.02$, respectively. On the other hand, the total logarithmic FAST-\hi mass of both galaxies is 7.97$\pm0.03$. Compared to the total MIGHTEE-\hi mass, the FAST-\hi excess fraction for this pair of galaxies is 32$\pm$6 percent, which is in line with the median value of 30 percent for the excess \hi in compact galaxy groups with the Green Bank Telescope (GBT) and the Very Large Array (VLA) data from \cite{Borthakur2010}. This excess of \hi gas is presumably outside the galaxies and may be considered to be in the CGM and IGM. We will return to this in the next section.

\begin{figure}
    \centering
    \includegraphics[width=0.9\columnwidth]{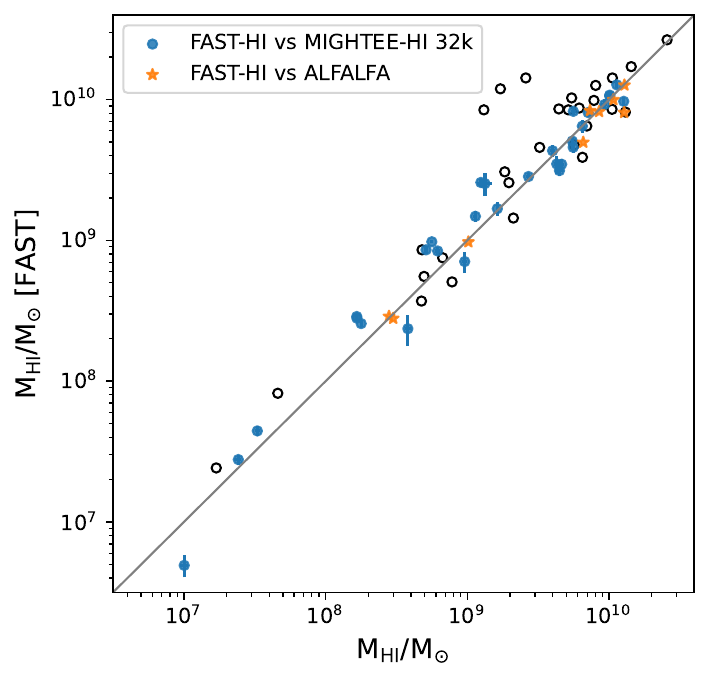}
  \caption{$M_{\rm HI}$ comparison of FAST-\hi  with MIGHTEE-\hi (blue dots) and ALFALFA detections (orange stars). The x-axis is the MIGHTEE-\hi  or ALFALFA \hi mass. The MIGHTEE-\hi data with the robustness of 0.5 are adopted for the mass comparison. The diagonal grey line is the one-to-one relation. Note that the black circles are the detections that have at least one close \hi or optical companion within the FAST beam, therefore their masses could be elevated due to multiple close companions.}
  \label{fig:masscomparison}
\end{figure}

\begin{table}
\caption{FAST-\hi COSMOS catalogue of isolated galaxies. Column 1: Source name; Column 2; FAST-\hi redshift with uncertainties less than 0.0001 estimated by the flux-weighted centroid position of source; Column 3: MIGHTEE-\hi redshift with uncertainties of $\sim$0.0005; Column 4: Optical spectroscopic redshift with the index indicating the reference; Column 5: FAST-\hi mass with 1$\sigma$ uncertainties; Column 6: Rest-frame velocity width $W_{50}$ in the unit of ${\rm km}\,{\rm s}^{-1}$. Note that the indexes of a, b, c, d, e, f, g and h for the optical redshifts are from \protect\cite{Polzin_2021}, \protect\cite{Ann_2015},  \protect\cite{Ahumada_2020}, \protect\cite{Davies_2018}, \protect\cite{Prescott_2006}, \protect\cite{Adame2024}, \protect\cite{Lilly_2007} and \protect\cite{Sohn_2019}.
}
\centering
\begin{tabular}{c@{\hskip 4pt}c@{\hskip 4pt}c@{\hskip 4pt}c@{\hskip 4pt}c@{\hskip 4pt}r}
\hline
\hline
\rule{0pt}{1.2em}
Source name (J2000) & \raisebox{0.1em}{$z_{\rm HI}^{\rm FAST}$} & \raisebox{0.1em}{$z_{\rm HI}^{\rm MGTH}$} & \raisebox{0.1em}{$z_{\rm opt}$} & $\log_{10}(M_{\rm HI})$  & $W_{50}$ \\
\hline
J100029.6+020853 & 0.0041  & 0.004 & 0.0041$^{\rm a}$ & 6.69$\pm$0.08 & 18.2 \\
J095846.1+022100 & 0.0058  & 0.006 & 0.0058$^{\rm b}$ & 8.46$\pm$0.01 & 108.5 \\
J100313.5+020600 & 0.0058  & 0.006 & 0.0058$^{\rm c}$ & 8.45$\pm$0.01 & 67.6 \\
J100226.9+021004 & 0.0060  & 0.006 & 0.0059$^{\rm d}$ & 7.44$\pm$0.04 & 33.1 \\
J095828.4+014133 & 0.0060  & 0.006 & 0.0060$^{\rm e}$ & 8.99$\pm$0.01 & 111.8 \\
J100006.0+015448 & 0.0063  & 0.006 & 0.0062$^{\rm d}$ & 7.65$\pm$0.04 & 67.4 \\
J095927.4+020042 & 0.0130  & 0.013 & 0.0130$^{\rm f}$ & 8.41$\pm$0.03 & 68.2 \\
J100221.0+022511 & 0.0214  & 0.021 & 0.0213$^{\rm f}$ & 8.93$\pm$0.03 & 94.6 \\
J100211.6+020114 & 0.0214  & 0.021 & 0.0213$^{\rm d}$ & 8.92$\pm$0.03 & 120.9 \\
J100001.5+023740 & 0.0243  & 0.024 & 0.0241$^{\rm g}$ & 8.37$\pm$0.11 & 74.5 \\
J095821.4+013509 & 0.0278  & 0.027 & 0.0278$^{\rm c}$ & 9.45$\pm$0.03 & 265.6 \\
J100113.8+021849 & 0.0286  & 0.028 & 0.0287$^{\rm d}$ & 9.41$\pm$0.03 & 158.2 \\
J095901.3+023334 & 0.0290  & 0.029 & 0.0288$^{\rm d}$ & 9.17$\pm$0.04 & 121.8 \\
J100313.6+022040 & 0.0389  & 0.039 & 0.0389$^{\rm f}$ & 9.64$\pm$0.04 & 239.4 \\
J100258.6+022045 & 0.0444  & 0.044 & 0.0442$^{\rm h}$ &10.10$\pm$0.02 & 277.0 \\
J100133.7+020656 & 0.0445  & 0.045 & 0.0442$^{\rm d}$ & 8.85$\pm$0.08 & 98.3 \\
J100054.6+022425 & 0.0445  & 0.044 & 0.0446$^{\rm d}$ & 9.70$\pm$0.03 & 258.7 \\
J100313.8+015344 & 0.0455  & 0.046 & 0.0455$^{\rm f}$ & 9.92$\pm$0.02 & 140.7 \\
J100142.6+024122 & 0.0470  & 0.047 & 0.0472$^{\rm c}$ & 9.91$\pm$0.02 & 192.6 \\
J100039.2+022459 & 0.0471  & 0.047 & 0.0470$^{\rm d}$ & 9.22$\pm$0.05 & 26.3 \\
J095921.9+024139 & 0.0476  & 0.048 & 0.0478$^{\rm d}$ & 9.66$\pm$0.04 & 255.3 \\
J100116.8+020315 & 0.0615  & 0.062 & 0.0615$^{\rm d}$ & 9.54$\pm$0.04 & 185.0 \\
J100149.2+022650 & 0.0622  & 0.062 & 0.0622$^{\rm d}$ & 9.40$\pm$0.08 & 151.1 \\
J100004.8+023836 & 0.0629  & 0.063 & 0.0631$^{\rm d}$ & 9.50$\pm$0.04 & 166.4 \\
J100003.5+015247 & 0.0650  & 0.065 & 0.0651$^{\rm d}$ & 9.81$\pm$0.05 & 200.4 \\
J095755.3+022542 & 0.0718  & 0.071 & 0.0721$^{\rm d}$ &10.03$\pm$0.04 & 338.7 \\
J100102.7+023111 & 0.0718  & 0.072 & 0.0718$^{\rm f}$ & 9.99$\pm$0.03 & 189.3 \\
J095755.3+022538 & 0.0718  & 0.071 & 0.0721$^{\rm d}$ &10.03$\pm$0.03 & 338.3 \\
J100106.1+015358 & 0.0776  & 0.078 & 0.0773$^{\rm d}$ & 9.96$\pm$0.04 & 199.7 \\
J095914.2+014906 & 0.0792  & 0.079 & 0.0791$^{\rm d}$ & 9.54$\pm$0.06 & 38.2 \\
\hline
\end{tabular}
\label{fast_cat}
\end{table}

We compare the \hi masses between the cross-matched FAST and MIGHTEE-\hi  detections and ALFALFA where available in Figure~\ref{fig:masscomparison}. Overall, the \hi masses from FAST detections  are in good agreement with those from the ALFALFA survey in orange dots, with median difference of $\sim$3 percent. The FAST-\hi detections tend to be systematically more \hi-massive than the MIGHTEE-\hi detections by a median fraction of $\sim7$ percent for the isolated galaxies, i.e. those with a single MIGHTEE catalogued source within the FAST beam, denoted by blue dots. This tendency is noticeable from the intermediate to high \hi mass end (i.e. $\geq$ $10^8$M$_{\odot}$), which is not unexpected as the FAST-\hi data are likely to pick up more faint diffuse \hi gas than the interferometric MIGHTEE-\hi observations. The origin of this limited diffuse gas could be due to the gas accretion from the large scale structure through the circumgalactic medium \citep{Hummels_2019,Sardone_2021}, and/or feedback from the galactic winds and fountains given that the galaxies in blue dots are relatively isolated \citep{Suresh_2015,Tumlinson_2017}. We present the FAST-\hi catalogue for isolated galaxies in Table~\ref{fast_cat}.

\begin{figure}
    \centering
    \includegraphics[width=\columnwidth]{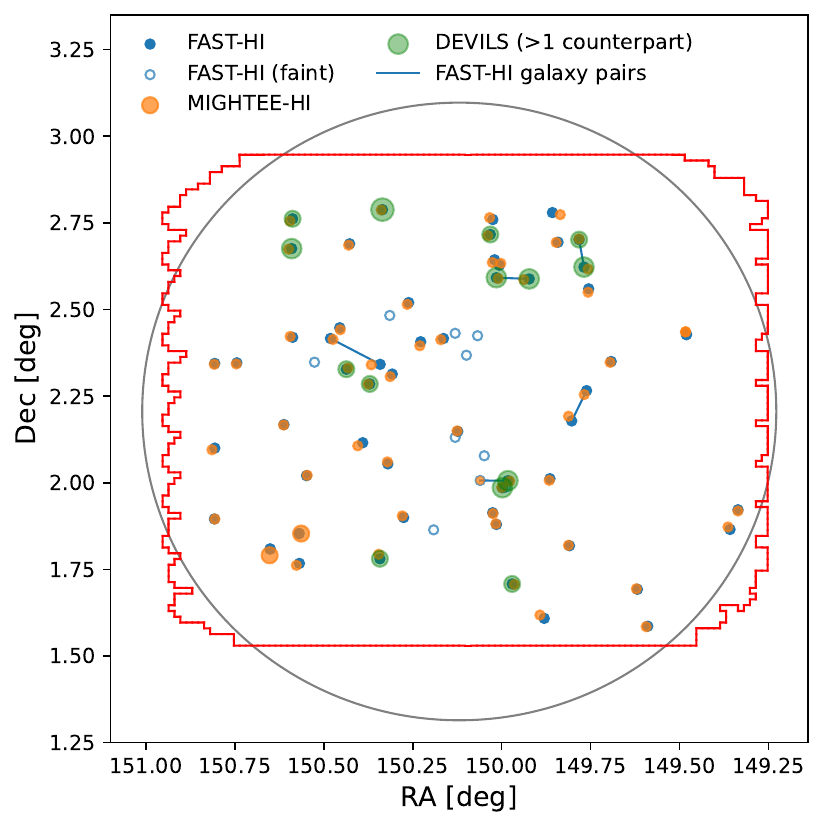}
  \caption{Angular distribution of the cross-matched FAST-\hi COSMOS sample with MIGHTEE-\hi and DEVILS surveys. The blue dots are the cross-matched FAST-\hi sources. The orange and green dots are the MIGHTEE-\hi and DEVILS counterparts within the FAST beam, respectively. The symbol size correlates with the number of the counterparts within the FAST beam, and we only show the DEVILS sample with at least two counterparts. The blue line connected dots are the FAST-\hi galaxy pairs bridged by a thin \hi structure. The blue circles are the faint FAST-\hi detections missed from our initial visual source finding but retrieved from the MIGHTEE-\hi positions. The red and black boundaries enclose the main surveyed areas from FAST-\hi and MIGHTEE-\hi respectively. }
  \label{fig:devils}
\end{figure}

The black circles in Figure~\ref{fig:masscomparison} are the \hi detections that have at least one additional close \hi or optical companion within the FAST beam, therefore their masses could be elevated  as their flux measurements cannot be well separated from their companions on the FAST-\hi images with a beam size of $\sim 3^\prime$. For example, the larger oranges dots (case 1) in Figure~\ref{fig:devils} are the cases where there are two MIGHTEE-\hi counterparts within a FAST beam, and the blue line connected dots (case 2) are the FAST-\hi galaxy pairs bridged by a thin \hi structure as shown for one of them in Figure~\ref{fig:example}. The green circles (case 3) symbolise at least two DEVILS spectroscopic counterparts within the FAST beam and relative frequency range corresponding to a velocity width of $\sim$200 ${\rm km}\,{\rm s}^{-1}$, and the symbol size correlates with the number of the optical counterparts. In cases 1 and 2, we add up all the MIGHTEE-\hi masses within the FAST beam or all the \hi masses in a galaxy pair to study the total \hi mass difference. We exclude the FAST-\hi sources covered by the green circles (case 3) in Figure~\ref{fig:devils} from the remaining cases to investigate \hi mass difference between FAST-\hi and MIGHTEE-\hi in the next section. However, they are included for studying the FAST-\hi mass-stellar mass relation where the DEVILS stellar masses within the FAST beam can be combined together.

\subsubsection{$M_{\rm HI}-M_{\star}$ relation}

\begin{figure}
    \centering
    \includegraphics[width=\columnwidth]{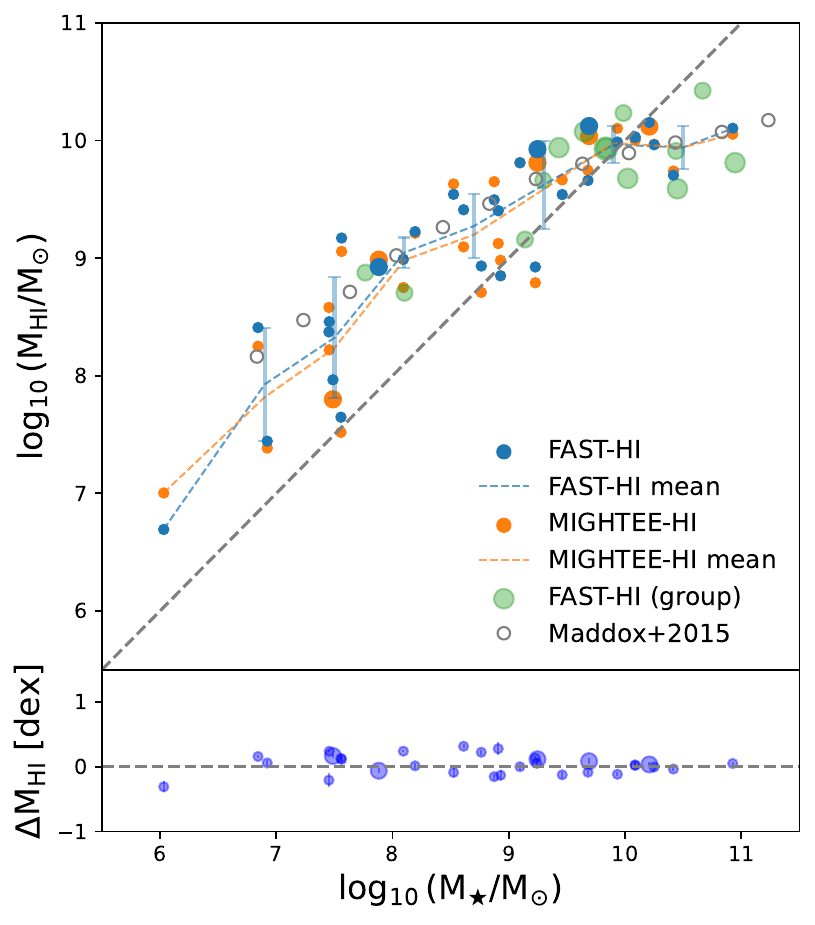}
  \caption{Top: $M_{\rm HI}-M_{\star}$ relation of the cross-matched FAST-\hi COSMOS sample with MIGHTEE-\hi. The blue and orange dots are from FAST-\hi and MIGHTEE-\hi samples respectively, while the green dots are the FAST-\hi galaxies with at least two DEVILS counterparts. The grey circles are the derived $M_{\rm HI}-M_{\star}$ relation from the ALFALFA galaxies by \protect\cite{Maddox2015}. Bottom: logarithmic \hi mass difference between the FAST-\hi and MIGHTEE-\hi samples as a function of the stellar mass. The symbol size correlates with the number of the MIGHTEE-\hi and DEVILS  counterparts within a FAST beam or in a galaxy pair in both panels.}
  \label{fig:mhi-mstar}
\end{figure}

The \hi and stellar mass ($M_{\rm HI}-M_{\star}$) relation is often used to investigate the processes of gas consumption and star formation in galaxies. In particular, exploring this relation indicates that there is an upper limit for \hi mass as a function of the stellar mass at high masses for \hi-selected samples \citep{huang2012,Maddox2015,Parkash2018,Pan_2023}. We show the $M_{\rm HI}-M_{\star}$ relation from our FAST-\hi sample in the upper panel of Figure~\ref{fig:mhi-mstar}. We bin the \hi sources in stellar mass with a bin width of 0.6 dex to estimate the mean \hi mass (blue dashed line) for isolated galaxies (small blue dots) and galaxy pairs (large blue dots). We find that the $M_{\rm HI}-M_{\star}$ relation from FAST-\hi is in good agreement with the relation from the ALFALFA-SDSS galaxy sample in \cite{Maddox2015} above the stellar mass of $\sim10^8$M$_\odot$, and detect a slope transition at $M_{\star}\sim10^{9.3}$M$_\odot$. Although our sample size of 35 is small after excluding a few case 3 sources, the FAST-\hi shows a higher \hi mass by a mean (or median) value of $\sim$0.04 dex (ranging up to $\sim$0.32 dex) when compared to the $M_{\rm HI}-M_{\star}$ relation from the MIGHTEE-\hi COSMOS sample, which is consistent with what we found in the one-to-one mass comparison in Figure~\ref{fig:masscomparison} as the number of case 1 and 2 galaxies are low. The FAST-\hi mass excess is most noticeable in the intermediate stellar masses which corresponds to intermediate and high \hi masses due to the non-linear relationship between \hi and stellar masses.

In the lower panel of Figure~\ref{fig:mhi-mstar}, we show the \hi mass difference between the FAST-\hi and MIGHTEE-\hi samples. The small dots are the isolated galaxies, whereas the large dots are case 1 and 2 galaxies which reside in a galaxy pair environment. Here we assume that the MIGHTEE-\hi gas is within the galaxies, hence we consider that the  \hi mass difference between MIGHTEE and FAST for isolated galaxies indicates the amount of the diffuse \hi gas in the CGM and/or IGM, and the mass difference between the FAST-\hi and MIGHTEE-\hi for galaxy pairs is indicative of the \hi gas in the CGM and IGM together.

It appears that the total \hi mass fraction in the IGM and CGM for the galaxy pairs is statistically higher than the \hi fraction for the isolated galaxies by $\sim13$ percent with the former being $\sim15\pm4$ percent and the latter being $\sim2\pm2$ percent. The median fractions of \hi gas surrounding the galaxy pairs and isolated galaxies are $18$ percent and 7 percent respectively, and the mass excess is weakly dependent on the stellar mass with slightly larger excess at the intermediate stellar masses based on our limited size of the \hi sample. This fraction difference is likely due to the contribution of the faint diffuse \hi gas in the IGM where the pristine gas is distributed as a reservoir or from a recent tidal interaction event and will possibly be accreted onto galaxies at a later stage \citep{Sancisi_2008,Wolfe_2013,Zhu2021,Zhou_2023}.

The green dots in the upper panel of Figure~\ref{fig:mhi-mstar} are likely small galaxy groups including potential galaxy pairs, and their FAST-\hi mass-stellar mass relation aligns with the average \hi mass-stellar mass relation of the rest \hi sample. The upcoming catalogue from source finding with the MIGHTEE 32k data should bring more information about the \hi gas distribution in the member galaxies of these groups, and help to determine the \hi gas fraction in the intragroup medium, which will shed light on the galaxy evolutionary stage.

\subsubsection{Sources of uncertainty}

\begin{figure*}
    \centering
\begin{subfigure}[s]{0.42\textwidth}
    \includegraphics[width=\columnwidth]{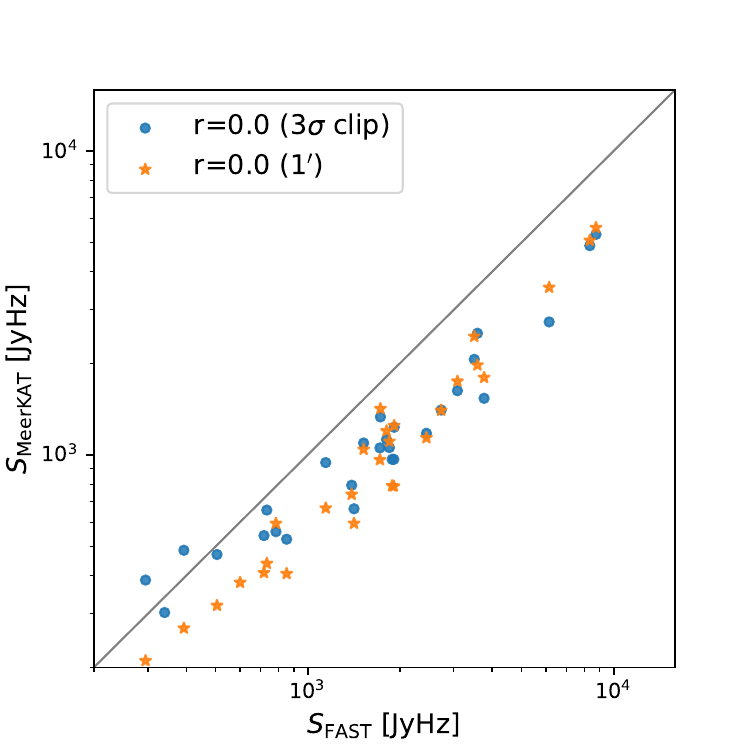}
\end{subfigure}
\begin{subfigure}[s]{0.42\textwidth}
    \includegraphics[width=\columnwidth]{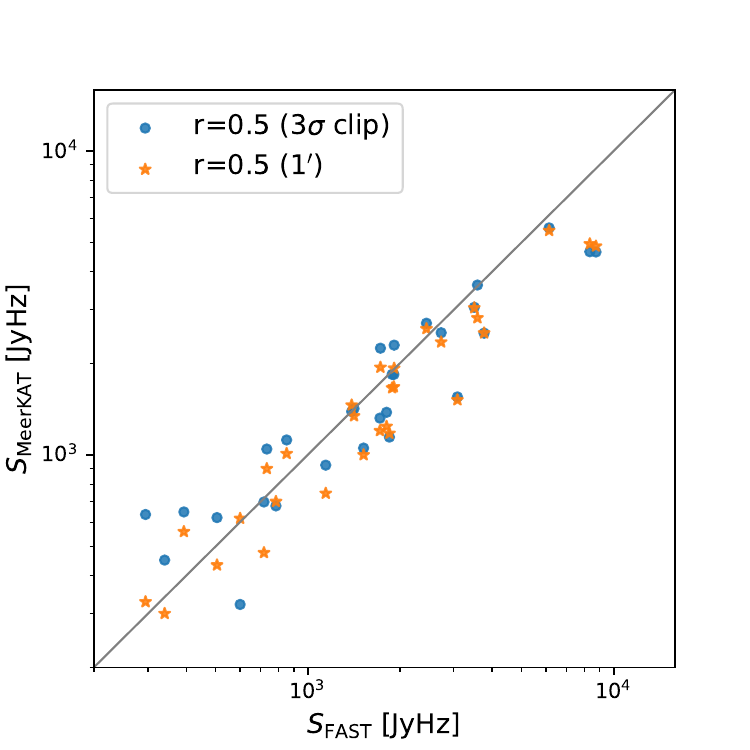}
\end{subfigure}
  \caption{\hi flux comparison between FAST and MeerKAT for isolated galaxies. The MIGHTEE-\hi robust-0.0 and 0.5 data against the FAST-\hi data are on the left and right panels respectively. For both panels, the blue dots and orange stars are the measurements with a 3$\sigma$ clipping approach and an extraction within a 1-arcminute aperture respectively. The diagonal grey lines are the one-to-one relations.}
  \label{fig:fast_vs_mgt_flux}
\end{figure*}

\begin{figure*}
    \centering
    \includegraphics[width=1.\textwidth]{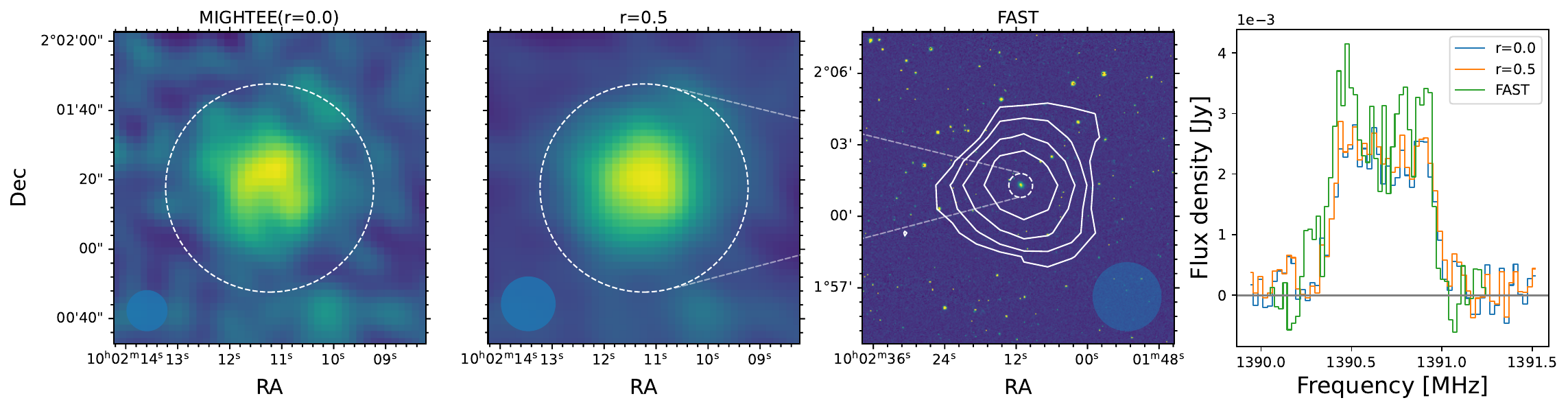}
  \caption{Example of extracting the FAST-\hi and MIGHTEE-\hi fluxes in a nearby spiral galaxy. From left to right panels, the MIGHTEE robust-0.0, robust-0.5, FAST-\hi moment-0 maps, and the flux density as a function of the frequency. The white dashed circles in the moment-0 maps indicate the 1$^\prime$ boundary and the white dashed lines are tracing the same angular size on different scales between the two middle panels. The blue dots in the corners are the corresponding beam sizes. The lowest column density contour on the FAST-\hi moment-0 map is about $6\times 10^{17}$\,$\mathrm{cm}^{-2}$ which is $\sim$3 times of the local noise level, and the contours increase by a factor of 2. The background optical image in g-band is from DECaLS.}
  \label{fig:fast_vs_mgt_flux_example}
\end{figure*}

The approach of measuring the diffuse \hi gas in CGM and IGM in this paper is to subtract the MIGHTEE-\hi gas from the total \hi content determined by the FAST-\hi gas, which relies on an assumption that the MIGHTEE-\hi measurement can well describe the \hi gas within the galaxies. To examine this assumption, we investigate the impact of two sources of uncertainty: 1) definition of \hi galaxy boundary, 2) missing flux due to the lack of zero spacing. We assume that the lack of additional field sources in the sky model for the primary calibrator used for the MIGHTEE-\hi observations does not significantly bias the resulting \hi mass measurements \citep[see e.g.][]{heywood2020}.

The \hi galaxy boundary is usually defined by exceeding a certain \hi column density, ranging from $10^{19}$ to $\sim10^{20}\,\mathrm{cm}^{-2}$. For example, \cite{Wang_2016} and \cite{Rajohnson_2022} use a column density of $1.25\times10^{20}\,\mathrm{cm}^{-2}$ that corresponds to a surface mass density of 1\,M$_\odot$ pc$^{-2}$ to determine the \hi disk diameter. However, there could be fainter or smaller \hi structures below that threshold but still associated with the galaxies, such as extraplanar \hi clouds and \hi tails \citep{Dedes_2010,Xu_2021}. \cite{Pingel_2018} and \cite{Sardone_2021} found that the cumulative \hi mass stays mostly flat below a critical column density of a few times $10^{19}\,\mathrm{cm}^{-2}$, and defined the diffuse neutral fraction as the fraction of \hi below column densities of $10^{19}\,\mathrm{cm}^{-2}$, which is of the same order as the sensitivity level of the MIGHTEE-\hi 32k data with r~=~0.5 over a 20 ${\rm km}\,{\rm s}^{-1}$ velocity width. Thus we adopt a 3$\sigma$ clipping approach to define the galaxy boundary, and compare the total flux within that boundary to the flux within a 1$^\prime$-diameter aperture (up to 2$^\prime$ for a few exceptions) in Figure~\ref{fig:fast_vs_mgt_flux}. The differences between the blue dots and orange stars are minimal for both the robust-0.0 and robust-0.5 data in the left and right panels, indicating that the 3$\sigma$ clipping approach is capable of recovering most of the \hi gas that is detectable by the MeerKAT.

In Figure~\ref{fig:fast_vs_mgt_flux_example}, we show an example of extracting the FAST-\hi and MIGHTEE-\hi fluxes in a spiral galaxy with a stellar mass of $1.7\times10^9$\,M$_{\odot}$. The FAST-\hi mass is $8.41\pm0.54\times10^8$\,M$_{\odot}$ which has a \hi mass excess of $\sim$27$\pm$5 percent compared to the MIGHTEE-\hi mass of $\sim6.17\pm0.23\times10^8$\,M$_{\odot}$. The MIGHTEE-\hi moment-0 map with the r~=~0.0 is shown in the left panel, while the MIGHTEE robust-0.5 moment-0 and the FAST-\hi maps are shown in the middle panels. The robust-0 image has the highest resolution of $\sim$12 arcseconds, and can therefore partially resolve this galaxy, featuring a faint \hi tail in the south-east region of the galaxy. The FAST-\hi excess as a function of frequency is clearly shown in the right panel where the MIGHTEE-\hi flux densities are extracted from a circular aperture indicated by the white dashed circles in the moment-0 maps with a diameter of 1 arcminute.

The impact of the zero spacing issue is strong for the robust-0.0 data across wide flux scales in the left panel of Figure~\ref{fig:fast_vs_mgt_flux}, but only appears to arise at the high flux end in the right panel, where the FAST-\hi fluxes are predominately larger than the MIGHTEE-\hi fluxes even for the robust-0.5 data with the maximum recoverable angular scale of $\sim$1 arcminute. This is unsurprising as galaxies with intermediate and high \hi masses dominate the high flux end while dwarf \hi galaxies tend to dominate the low flux end and lack the ability of hosting a large fraction of \hi gas in their circumgalatic medium, therefore there is no strong FAST-\hi flux excess for the dwarfs. As some galaxies in our sample are larger than 1 arcminute, we note that there could be a small fraction of flux missing from the MIGHTEE-\hi robust-0.5 data when the \hi masses within the galaxies are determined. However, considering that MeerKAT's antennas are densely packed and the number of galaxies larger than 1 arcminute are not dominating our sample, the impact of the zero spacing issue on the flux measurements within the galaxy disks for the robust-0.5 data is limited. 

\subsection{FAST-MeerKAT synergy}

With the $\sim 3^\prime$ beam size of FAST, we can at least partially resolve some galaxy groups and pairs, or even a few large nearby galaxies in our data \citep[e.g.][]{Wang2023,Lin2023}. By convolving the MIGHTEE-\hi data inside these group and pair members or large nearby galaxies to the same angular resolution of FAST, we are able to subtract the individual galaxy contributions from the FAST intensity map, leaving the residual that is the diffuse \hi gas distributed in IGM/CGM resolved out by MeerKAT. For dwarf galaxies at low redshifts and galaxy groups at high redshifts ($z>0.2$), most of the targets cannot be resolved by FAST. We will therefore concentrate efforts on measuring the total amount of \hi gas in IGM/CGM within various environments (e.g. isolated galaxies or galaxy groups) by subtracting the MIGHTEE-\hi fluxes from the FAST-\hi images.

We are also able to cross-correlate the FAST \hi intensity maps with optical galaxy catalogues \citep[e.g.][]{chang2010hydrogen,Tramonte_2020,Wolz2022} in order to mitigate foreground and systematic effects and thus help to constrain cosmological parameters such as the total \hi mass density ($\Omega_{\rm HI}$), the dark matter power spectrum and their evolution. Moreover, the cross-correlation with the MIGHTEE-\hi intensity map from interferometric observations can give a high-SNR detection of the large scale \hi structure due to the exceptional sensitivity of both radio instruments, and provide a complementary view of the \hi mass function from the \hi detections alone. 

\section{Conclusions}
\label{sec:conclusions}

The FAST-\hi COSMOS survey presented in this paper demonstrates the capability of the FAST telescope in detecting the faint \hi signals. We describe the data processing of the FAST raw data across the frequency band of 1310-1420 MHz, showing that the FAST data have reached a median 3$\sigma$ column density of $N_{\rm{HI}} \sim 2\times 10^{17}$ cm$^{-2}$ over a 5 ${\rm km}\,{\rm s}^{-1}$ channel width. This sensitivity allows us to study the distribution of \hi gas within galaxies, CGM and IGM when combined with the high-resolution of the MeerKAT telescope.

We searched the \hi spectral cube and visually identified  a total of 80 sources, of which $\sim$70 percent are cross-matched with the MIGHTEE-\hi catalogue. We examined the \hi mass against redshift and the \hi flux against the velocity width $W_{50}$, and we found that the non cross-matched sample is dominated by the low-flux or narrow-line width \hi galaxies. We compared the \hi masses between the FAST-\hi and MIGHTEE-\hi 32k data and find that the FAST-\hi masses are systematically higher than the MIGHTEE-\hi masses by a median fraction of $\sim7$\% for isolated galaxies alone.

We further studied the $M_{\rm HI}-M_{\star}$ relation in the last billion years with the cross-matched sources, and find that the FAST-\hi sources exhibit higher \hi masses on average by a mean value of $\sim$0.04 dex (ranging up to 0.32 dex) for the isolated galaxies and galaxy pairs when compared to the \hi masses from the MIGHTEE-\hi COSMOS sample. We find that the major FAST-\hi mass excess is contributed by galaxies with intermediate stellar masses based on our relatively small sample size.

By separating the galaxy pairs from the isolated galaxies, we find the total \hi gas fraction in the IGM and CGM together in the galaxy pairs is statistically higher than the gas fraction in the CGM (and/or IGM) of isolated galaxies by $\sim13$\%. The former being $\sim15\pm4$\% and the latter being $\sim2\pm2$\% suggest that the CGM and IGM associated with interacting galactic systems are \hi gas richer than those surrounding the isolated galaxies (possibly due to the gas accretion from the nearby reservoir or tidal interaction between the pair galaxies), albeit with large uncertainties in the measurements as a result of the small sample size of the galaxies. Taken together, the diffuse \hi in the CGM and the IGM accounts for $\sim5\pm2$\% of the total \hi gas in relatively simple galactic environments (i.e. without considering the complex galaxy groups).

By combining the capabilities of FAST and MeerKAT telescopes, the full potential of both instruments can be explored. We highlight the prospects of detecting the faint diffuse \hi gas in IGM and CGM, and studying the evolution of the $\Omega_{\rm HI}$. This FAST-\hi COSMOS survey acts as a pilot project to form a foundation for further collaboration between MeerKAT and FAST on various astrophysical and cosmological applications with much wider area surveys.

\section*{Data availability}
The FAST-\hi COSMOS spectral data are available upon reasonable request. All the raw data of the FAST observations will be freely accessible to any user after a twelve-month privileged period as per the FAST data release policy \footnote{\url{https://fast.bao.ac.cn/cms/article/129}}.

\section*{Acknowledgements}
We thank the anonymous referee for their suggestions that have improved this paper. We thank Di Li, Bradley Frank, Chun Sun, Tobias Westmeier and Hongwei Xi for useful discussions. This work has used the data from the Five-hundred-meter Aperture Spherical radio Telescope (FAST).  FAST is a Chinese national mega-science facility, operated by the National Astronomical Observatories of Chinese Academy of Sciences (NAOC). 

The MeerKAT telescope is operated by the South African Radio Astronomy Observatory, which is a facility of the National Research Foundation, an agency of the Department of Science and Innovation. We acknowledge use of the IDIA data intensive research cloud for data processing. The IDIA is a South African university partnership involving the University of Cape Town, the University of Pretoria and the University of the Western Cape. The authors acknowledge the Centre for High Performance Computing (CHPC), South Africa, for providing computational resources to this research project.

We acknowledge the use of the Ilifu cloud computing facility - \url{www.ilifu.ac.za}, a partnership between the University of Cape Town, the University of the Western Cape, the University of Stellenbosch, Sol Plaatje University, the Cape Peninsula University of Technology and the South African Radio Astronomy Observatory. The Ilifu facility is supported by contributions from IDIA and the Computational Biology division at UCT and the Data Intensive Research Initiative of South Africa (DIRISA).

HP, MJJ and IH acknowledge support from a UKRI Frontiers Research Grant [EP/X026639/1], which was selected by the European Research Council. HP, MJJ, MGS and IH acknowledge support from the South African Radio Astronomy Observatory which is a facility of the National Research Foundation (NRF), an agency of the Department of Science and Innovation.  
MJJ acknowledge generous support from the Hintze
Family Charitable Foundation through the Oxford Hintze Centre
for Astrophysical Surveys and the UK Science and Technology Facilities Council [ST/S000488/1]. HP also acknowledges the warm hospitality at the Purple Mountain Observatory in 2021. YZM is supported by the National Research Foundation of South Africa under Grants No. 150580, No. ERC23040389081, No. CHN22111069370 and No. RA211125652169. MGS also acknowledges support from the National Research Foundation (Grant No. 84156). AAP, MJJ and IH acknowledge support of the STFC consolidated grant [ST/S000488/1] and  [ST/W000903/1]. YC thanks the Center for Astronomical Mega-Science, Chinese Academy of Sciences, for the FAST distinguished young researcher fellowship (19-FAST-02). YC also acknowledges the support from the National Natural Science Foundation of China (NSFC) under grant No. 12050410259 and the Ministry of Science and Technology (MOST) of China grant no. QNJ2021061003L. YPJ is supported by NSFC (12133006) and by 111 project No. B20019.

\bibliographystyle{mnras}
\bibliography{references}




\appendix

\section{Injection and recovery}
\label{sec:injection}

Since the noise in our FAST-\hi data cube does not follow a perfect Gaussian distribution (as shown in Figure~\ref{fig:rmsvsfreq}) due to complex systematic effects present in our data, we use fake sources with known \hi masses to verify that optimal input parameters are employed by SoFiA-2 on our FAST data.

We first replace the visually-identified sources in the \hi spectral cube with noise from their neighboring channels, and inject fake \hi sources with similar \hi masses and velocity widths of the real detections to random positions centered at $z_{\rm HI}$ with a deviation of 0.001. We then repeat the injection 10 times for each detection to avoid low-number statistics. For each iteration, we randomly generate the flux density profile, following a form of the busy function \citep{Westmeier_2013,Pan2020}.

\begin{figure*}
    \centering
\begin{subfigure}[s]{0.33\textwidth}
    \includegraphics[width=\columnwidth]{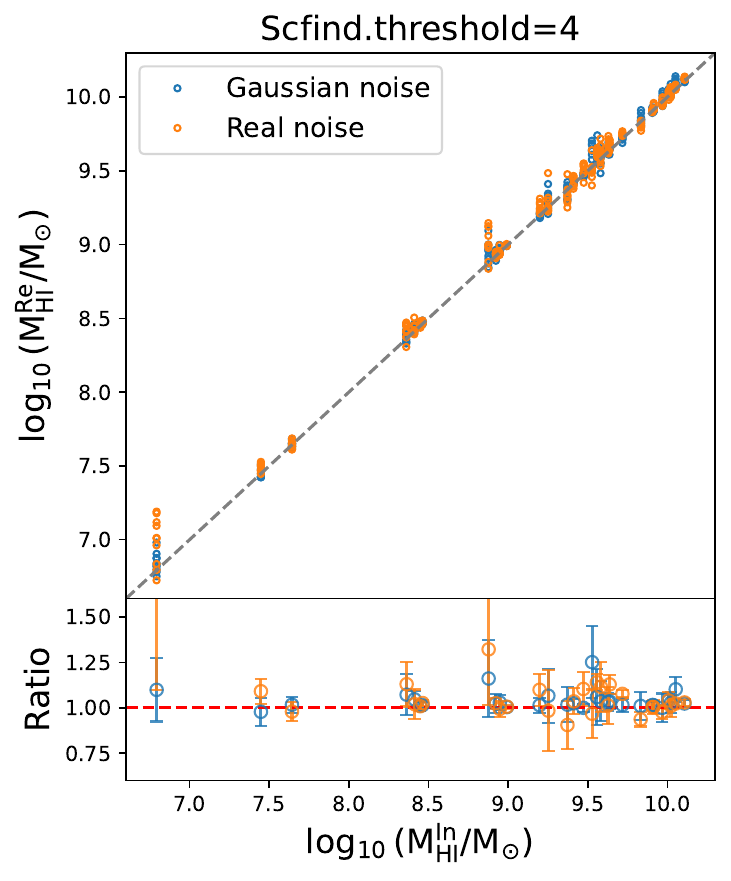}
\end{subfigure}
\begin{subfigure}[s]{0.33\textwidth}
    \includegraphics[width=\columnwidth]{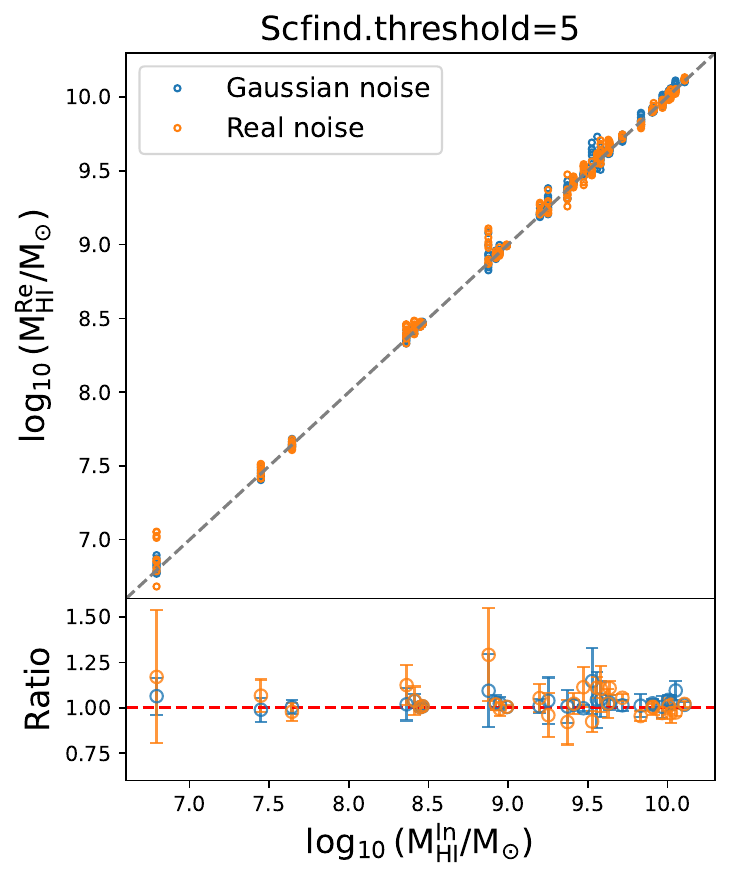}
\end{subfigure}
\begin{subfigure}[s]{0.33\textwidth}
    \includegraphics[width=\columnwidth]{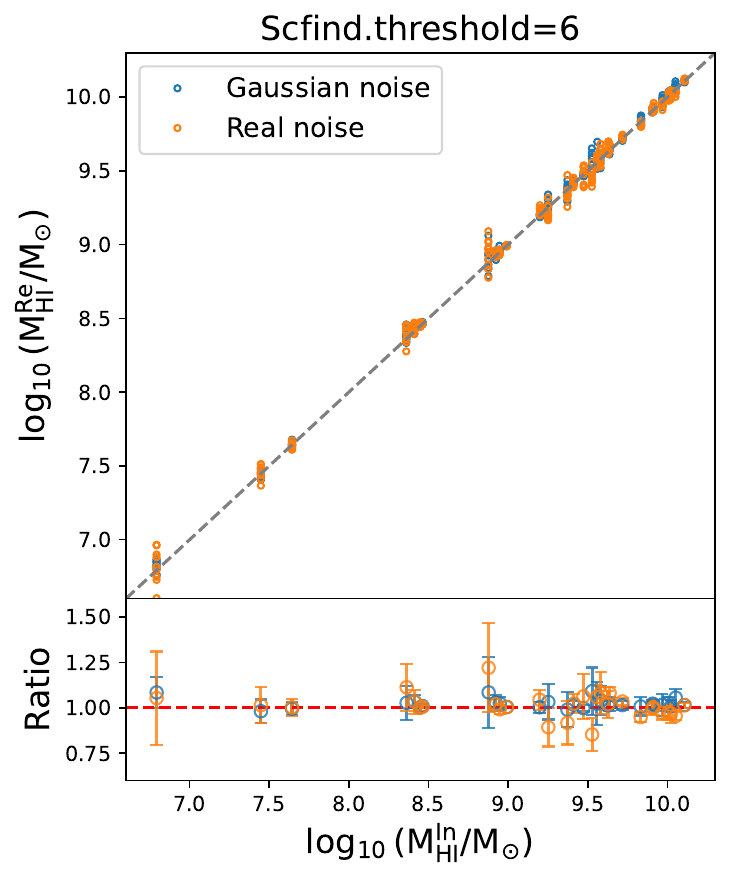}
\end{subfigure}
  \caption{Top: \hi mass comparison between injected and recovered sources. The \hi masses are estimated by employing the SoFiA-2 with the flux threshold of \textit{scfind.threshold}=[4, 5, 6] from left to right panels,  respectively. The fake sources are injected to \hi cubes with Gaussian noise in blue and Real noise in orange where the genuine sources are removed. The diagonal grey dashed lines are the one-to-one relations. Bottom: median ratio of the recovered \hi mass $M_{\rm HI}^{\rm Re}$ to the injected \hi mass $M_{\rm HI}^{\rm In}$ with 1$\sigma$ uncertainties as a function of the injected \hi mass. }
  \label{fig:input_recovery}
\end{figure*}

We show the recovered \hi mass $M_{\rm HI}^{\rm Re}$ against the injected \hi mass $M_{\rm HI}^{\rm In}$ in the top panel of Figure~\ref{fig:input_recovery} with three flux thresholds, i.e. \textit{scfind.threshold}=[4, 5, 6], used by the Smooth+Clip finder of SoFiA-2. The median ratio of $M_{\rm HI}^{\rm Re}$ to $M_{\rm HI}^{\rm In}$ as a function of the injected \hi mass is shown in the bottom panel. As our sample is dominated by the bright \hi sources, the observed median mass ratios of [1.026, 1.01, 0.994] are very close to 1, in a decreasing order for the increasing threshold from the left to right panels as expected. We also model a Gaussian background noise in comparison with the real noise, and find that the recovered mass ratio follows a similar trend to the case of real noise but with smaller uncertainties due to the well-behaved fake noise.

\begin{figure}
    \centering
    \includegraphics[width=\columnwidth]{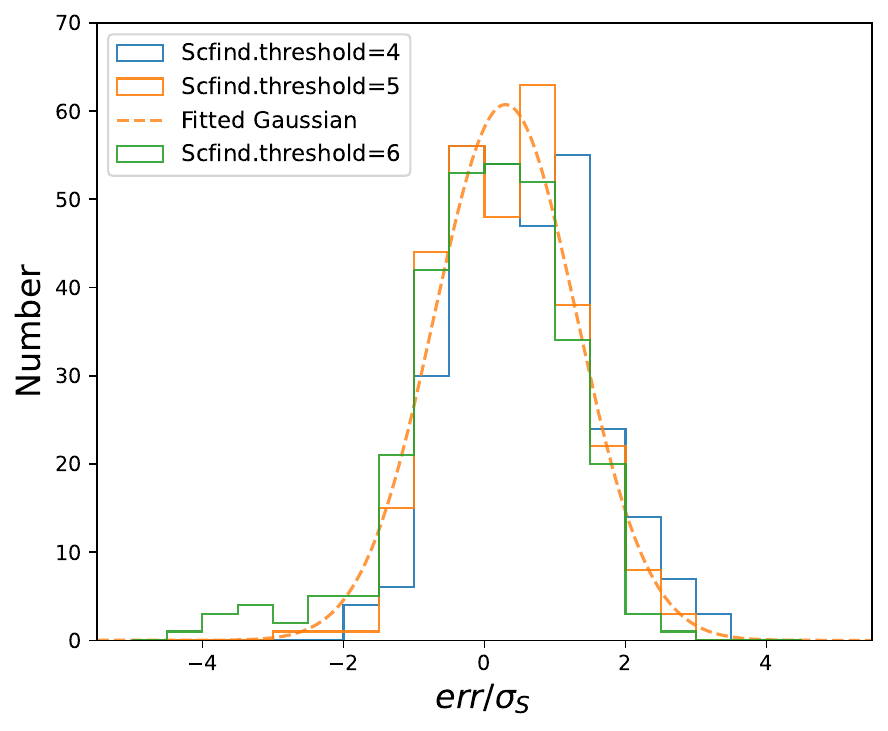}
  \caption{Histogram of flux measurement error ($err$) divided by the flux measurement uncertainty ($\sigma_s$) derived by SoFiA-2 for the mock galaxies. The dashed, orange line
shows the result of a Gaussian fit to the histogram with a peak position of 0.3 ± 0.1 and a standard deviation of 1.01 ± 0.06, when we chose the best threshold \textit{scfind.threshold}=5 for the FAST-\hi flux measurements.}
  \label{fig:error}
\end{figure}

To further examinate the accuracy of the recovered fluxes and their associated uncertainties, we show a histogram of the flux measurement error ($err=S^{\rm Re}-S^{\rm In}$) divided by the flux measurement uncertainty ($\sigma_s$) from SoFiA-2. Under ideal conditions, $err/\sigma_s$ should have a Gaussian distribution with a centroid of 0 and a standard deviation of 1. A Gaussian fit to the histogram with \textit{scfind.threshold}=5 yields a centroid of 0.3 ± 0.1 and a standard deviation of 1.01 ± 0.06, which is close to the expected values and indicates that the flux uncertainties reported by SoFiA-2 are accurate as demonstrated by \cite{Westmeier_2021}. The small offset agrees with the median ratio of 1.01 as we see in the middle panel of Figure~\ref{fig:input_recovery}. When the threshold is increased to 6, the offset decreases slightly, as the green histogram shows in Figure~\ref{fig:error}. There appears to be a positive flux bias of 2–3 percent from SoFiA-2, which is assumed to be related to a default parameter of \textit{scfind.replacement} as reported by \cite{Westmeier_2021}. Considering that the returned uncertainties are far larger than this small positive flux bias and we would like to retain a certain sensitivity to the diffuse \hi gas surrounding the galaxies, we chose the \textit{scfind.threshold}=5 as the best parameter to measure the total fluxes. We note that a choice of slightly higher threshold will not change our main results.

\section{Non-detections}
\label{sec:non-det}

\begin{figure*}
    \centering
\begin{subfigure}[t]{0.32\textwidth}
    \includegraphics[width=\columnwidth]{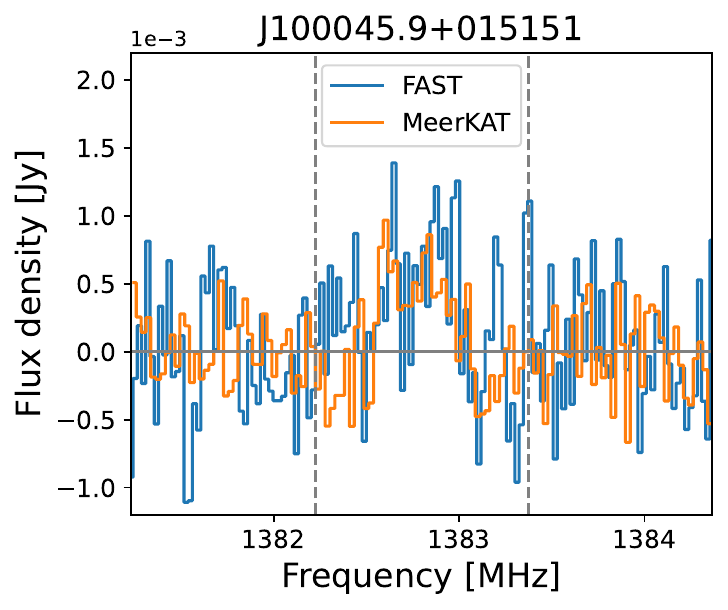}
    \includegraphics[width=\columnwidth]{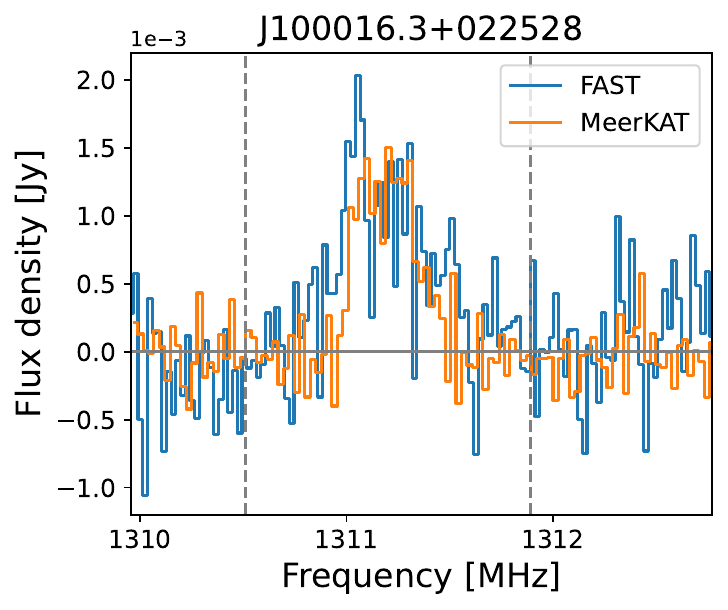}
    \includegraphics[width=\columnwidth]{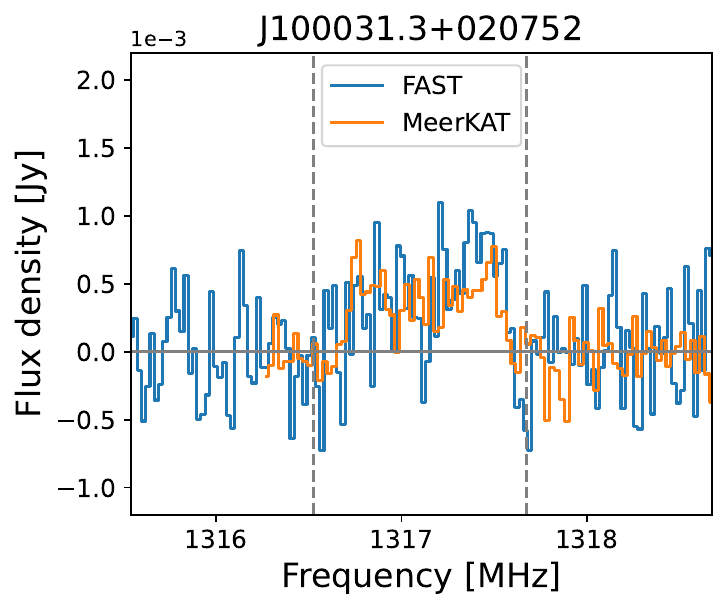}
\end{subfigure}
\hspace{-2.5mm}
\begin{subfigure}[t]{0.32\textwidth}
    \includegraphics[width=\columnwidth]{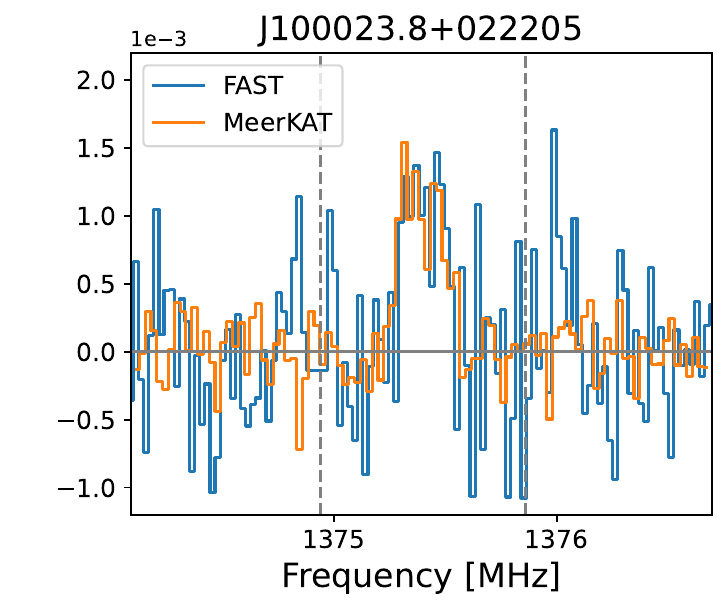}
    \includegraphics[width=\columnwidth]{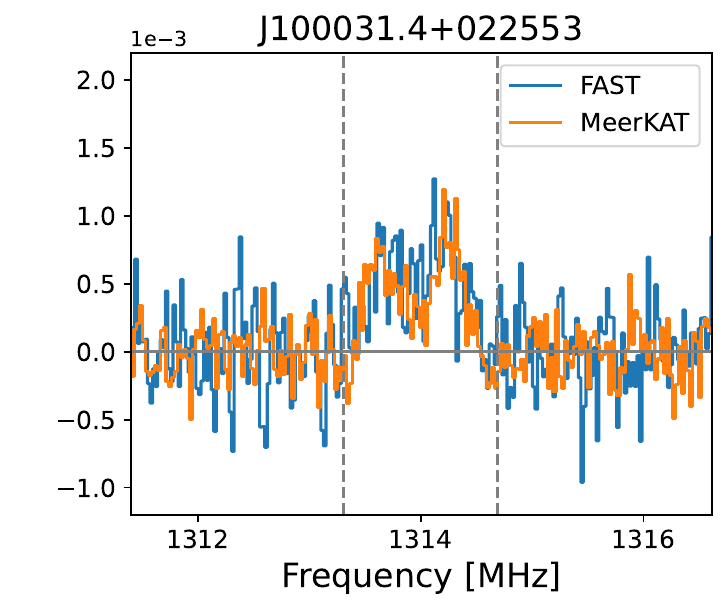}
    \includegraphics[width=\columnwidth]{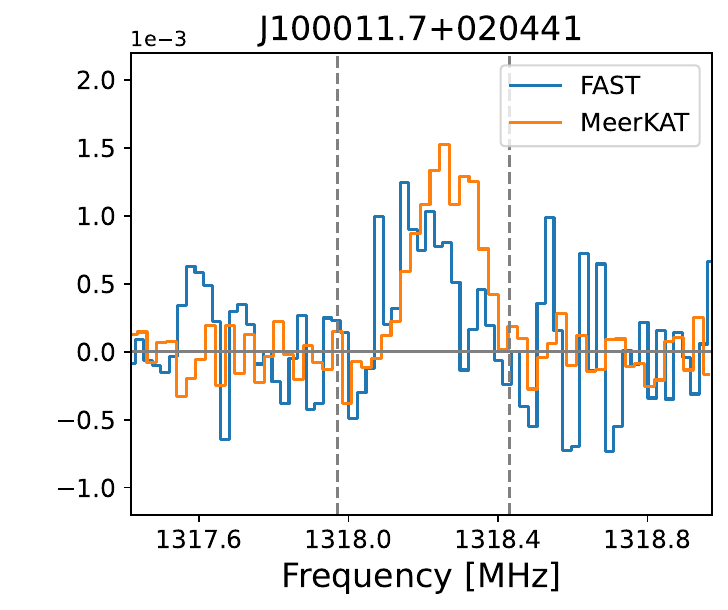}
\end{subfigure}
\hspace{-2.5mm}
\begin{subfigure}[t]{0.32\textwidth}
    \includegraphics[width=\columnwidth]{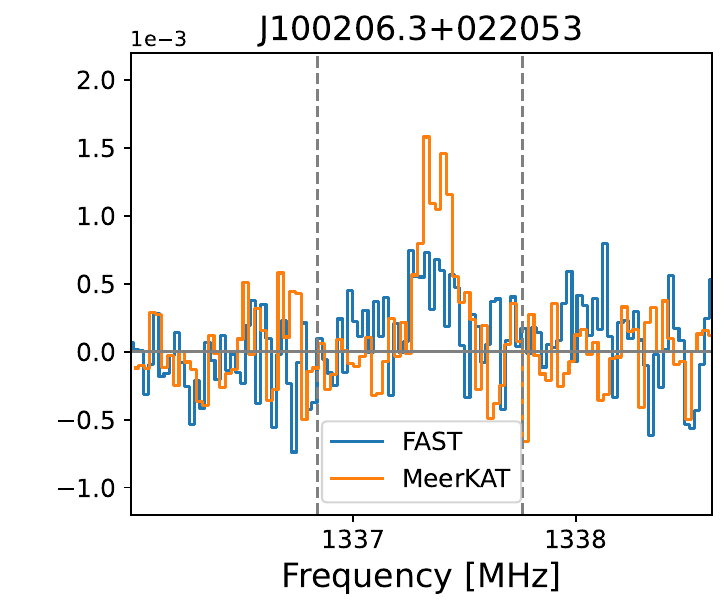}
    \includegraphics[width=\columnwidth]{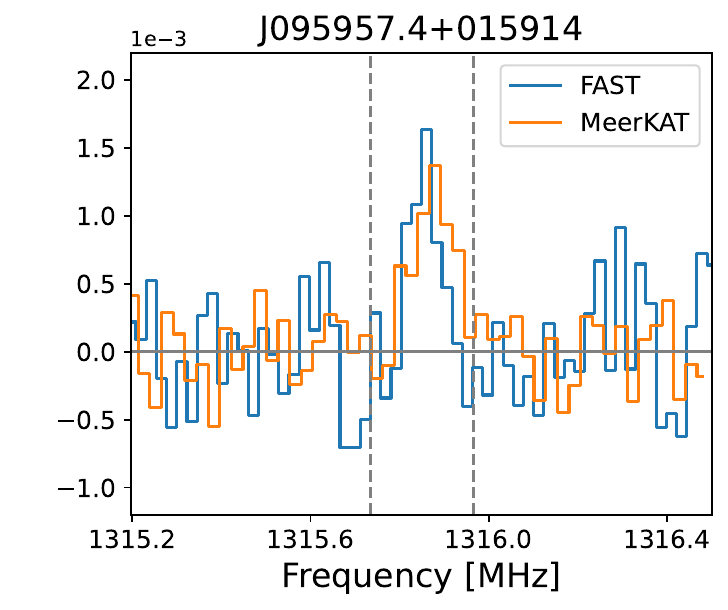}
    \includegraphics[width=\columnwidth]{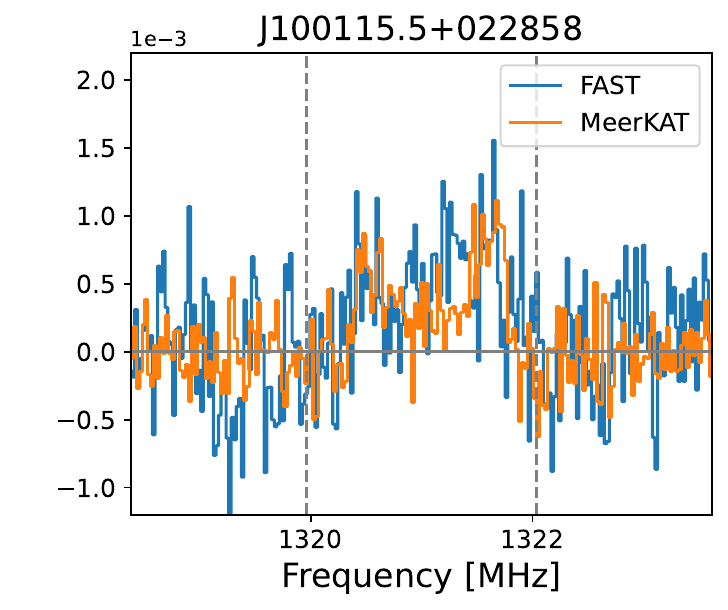}
\end{subfigure}

  \caption{Comparison of the \hi flux density profiles between FAST and MeerKAT for 9 galaxies that are not visually identified in the FAST data. The grey dashed lines indicate the spectral limits used for calculating the integrated fluxes.}
  \label{fig:non-det}
\end{figure*}

\begin{figure}
    \centering
    \includegraphics[width=0.8\columnwidth]{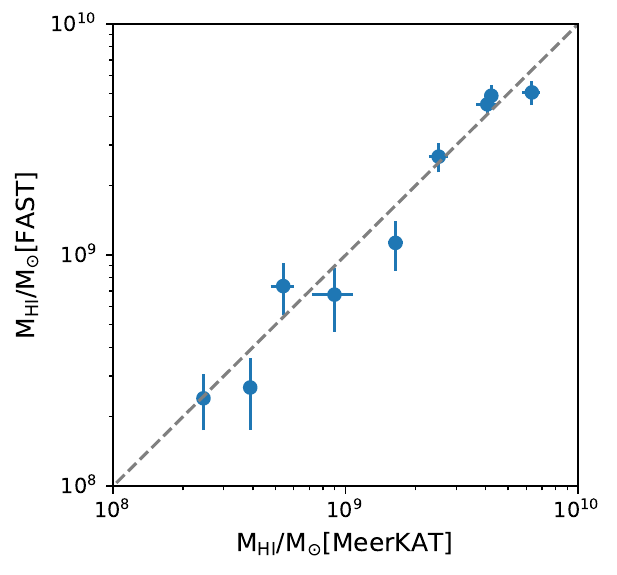}
  \caption{\hi mass comparison between FAST and MeerKAT for 9 galaxies that are not visually identified in the FAST data. The diagonal grey dashed line is the one-to-one relation.}
  \label{fig:non-dec-mass}
\end{figure}

We cross-matched the FAST-\hi and MIGHTEE-\hi detections in the COSMOS field, and found that there are 9 sources that are present in MIGHTEE-\hi catalogue but not visually identified in our FAST data. To examine the cause, we extract the fluxes from the FAST-\hi cube at the location of MIGHTEE-\hi detections, and compare their flux density profiles with those from MeerKAT data in Figure~\ref{fig:non-det} and their \hi masses in Figure~\ref{fig:non-dec-mass}, indicating that the FAST-\hi fluxes are largely in good agreement with the MeerKAT fluxes. The fact that these sources did not pass our visual filtering is due to their low signal to noise ratios and/or narrower velocity widths, and we were doing visual source finding on a coarser \hi cube with a velocity resolution of 26 ${\rm km}\,{\rm s}^{-1}$. Although these faint sources show up when we zoom in the data cube, we decide to list them in Table~\ref{tab:fast_cat_non} as non-detections, where the FAST-\hi masses are estimated by integrating the flux densities within the vertical grey dashed lines in Figure~\ref{fig:non-det}.

\begin{table}
\caption{FAST-\hi COSMOS catalogue of non-detections. The FAST-\hi redshifts and masses are displayed in the second and third columns with the signal to noise ratio (SNR) in the fourth column.}
\label{tab:fast_cat_non}
\centering
\begin{tabular}{ccccr}
\hline
\hline
Source name (J2000) & $z_{\rm HI}$ & $\log_{10}(M_{\rm HI})$ & SNR \\
\hline
J100045.9+015151 & 0.0272 & 8.38 & 3.8 \\
J100023.8+022205 & 0.0327 & 8.43 & 3.0 \\
J100206.3+022053 & 0.0621 & 8.87 & 4.1 \\
J100115.5+022858 & 0.0753 & 9.71 & 8.7 \\
J100011.7+020441 & 0.0775 & 9.05 & 4.2 \\
J100031.3+020752 & 0.0784 & 9.43 & 7.1 \\
J095957.4+015914 & 0.0795 & 8.83 & 3.3 \\
J100031.4+022553 & 0.0810 & 9.65 & 11.6 \\
J100016.3+022528 & 0.0833 & 9.69 & 8.9 \\
\hline
\end{tabular}
\end{table}


\end{document}